\documentclass[journal]{IEEEtran}
\usepackage{amsmath,amsfonts,amssymb}
\usepackage[ruled,linesnumbered]{algorithm2e}
\usepackage{array}
\usepackage[caption=false,font=normalsize,labelfont=sf,textfont=sf]{subfig}
\usepackage{textcomp}
\usepackage{stfloats}
\usepackage{url}
\usepackage{verbatim}
\usepackage{graphicx}
\usepackage{cite}
\usepackage{ntheorem}
\usepackage{booktabs}
\usepackage{hyperref}
\usepackage{tikz,xcolor,hyperref}% Make Orcid icon
\definecolor{lime}{HTML}{A6CE39}
\DeclareRobustCommand{\orcidicon}{%
    \begin{tikzpicture}
    \draw[lime, fill=lime] (0,0) 
    circle [radius=0.16] 
    node[white] {{\fontfamily{qag}\selectfont \tiny ID}};    \draw[white, fill=white] (-0.0625,0.095) 
    circle [radius=0.007];    \end{tikzpicture}
    \hspace{-2mm}}
\foreach \x in {A, ..., Z}{%
    \expandafter\xdef\csname orcid\x\endcsname{\noexpand\href{https://orcid.org/\csname orcidauthor\x\endcsname}{\noexpand\orcidicon}}
    }

\theoremseparator{.}
\newskip\theorempreskipamount
\newskip\theorempostskipamount

\theorembodyfont{}

\begin{document}

\title{Low-Altitude Satellite-AAV Collaborative Joint Mobile Edge Computing and Data Collection via Diffusion-based Deep Reinforcement Learning}

\author{Boxiong Wang\orcidA{}, 
Hui Kang\orcidB{}, 
Jiahui Li\orcidC{}, 
Geng Sun\orcidD{}, \IEEEmembership{Senior Member, IEEE,}
Zemin Sun\orcidE{}, 
Jiacheng Wang\orcidF{},
Dusit Niyato\orcidG{},~\IEEEmembership{Fellow, IEEE},
and Shiwen Mao\orcidH{},~\IEEEmembership{Fellow, IEEE}
        % <-this % stops a space
\thanks{This work is supported in part by the National Natural Science Foundation of China (62272194, 62471200), in part by the Science and Technology Development Plan Project of Jilin Province (20250102210JC, 20250101027JJ), in part by the Seatrium New Energy Laboratory, Singapore Ministry of Education (MOE) Tier 1 (RT5/23 and RG24/24), the Nanyang Technological University (NTU) Centre for Computational Technologies in Finance (NTU-CCTF), and the Research Innovation and Enterprise (RIE) 2025 Industry Alignment Fund - Industry Collaboration Projects (IAF-ICP) (Award I2301E0026), administered by Agency for Science, Technology and Research (A*STAR), in part by the Postdoctoral Fellowship Program of China Postdoctoral Science Foundation (GZC20240592), in part by the China Postdoctoral Science Foundation General Fund (2024M761123), in part by the Scientific Research Project of Jilin Provincial Department of Education (JJKH20250117KJ), and in part by the Graduate Innovation Fund of Jilin University (2025CX210). \textit{(Corresponding authors: Jiahui Li and Geng Sun.)}
\par Boxiong Wang is  with the College of Computer Science and Technology, Jilin University, Changchun 130012, China, and also with Beijing Zhongguancun Academy, Beijing 100089, China. (Email: wangbx0320@163.com).
\par Jiahui Li and Zemin Sun are with the College of Computer Science and Technology, Jilin University, Changchun 130012, China. (E-mails: lijiahui@jlu.edu.cn; sunzemin@jlu.edu.cn).
\par Hui Kang is with the College of Computer Science and Technology, Jilin University, Changchun 130012, China, and also with the Key Laboratory of Symbolic Computation and Knowledge Engineering of Ministry of Education, Jilin University, Changchun 130012, China. (E-mail: kanghui@jlu.edu.cn).
\par Geng Sun is with the College of Computer Science and Technology, Jilin University, Changchun 130012, China, and also with the Key Laboratory of Symbolic Computation and Knowledge Engineering of Ministry of Education, Jilin University, Changchun 130012, China. He is also with the College of Computing and Data Science, Nanyang Technological University, Singapore 639798 (E-mail: sungeng@jlu.edu.cn).
\par Jiacheng Wang and Dusit Niyato are with the College of Computing and Data Science, Nanyang Technological University, Singapore (E-mails: jiacheng.wang@ntu.edu.sg; dniyato@ntu.edu.sg).
\par Shiwen Mao is with the Department of Electrical and Computer Engineering, Auburn University, Auburn, AL 36849-5201 USA (E-mail: smao@ieee.org).
}
}

\maketitle

\begin{abstract}
The integration of satellite and autonomous aerial vehicle (AAV) communications has become essential for the scenarios requiring both wide coverage and rapid deployment, particularly in remote or disaster-stricken areas where the terrestrial infrastructure is unavailable. Furthermore, emerging applications increasingly demand simultaneous mobile edge computing (MEC) and data collection (DC) capabilities within the same aerial network. However, jointly optimizing these operations in heterogeneous satellite-AAV systems presents significant challenges due to limited on-board resources and competing demands under dynamic channel conditions. In this work, we investigate a satellite-AAV-enabled joint MEC-DC system where these platforms collaborate to serve ground devices (GDs). Specifically, we formulate a joint optimization problem to minimize the average MEC end-to-end delay and AAV energy consumption while maximizing the collected data. Since the formulated optimization problem is a non-convex mixed-integer nonlinear programming (MINLP) problem, we propose a Q-weighted variational policy optimization-based joint AAV movement control, GD association, offloading decision, and bandwidth allocation (QAGOB) approach. Specifically, we reformulate the optimization problem as an action space-transformed Markov decision process to adapt the variable action dimensions and hybrid action space. Subsequently, QAGOB leverages the multi-modal generation capacities of diffusion models to optimize policies and can achieve better sample efficiency while controlling the diffusion costs during training. Simulation results show that QAGOB outperforms five other benchmarks, including traditional DRL and diffusion-based DRL algorithms. Furthermore, the MEC-DC joint optimization achieves significant advantages when compared to the separate optimization of MEC and DC.

\end{abstract}


%
%%%%%%%%%%%%%%%%%%%%%% Introduction %%%%%%%%%%%%%%%%%%%%%%%%%%%%%%%%%
%

\section{Introduction}
\label{sec:Introduction}

\par \IEEEPARstart{T}{he} advancements in wireless communications and the development of manufacturing technologies have driven an exponential growth of Internet of Things (IoT) applications. These applications simultaneously require ubiquitous network coverage, robust distributed data storage, and real-time computing service capabilities. However, due to the limited computing and storage resources, IoT ground devices (GDs) are typically unable to effectively handle the compute-intensive tasks and store the large amounts of data by themselves. In this context, mobile edge computing (MEC) has emerged as a paradigm for computational offloading, while data collection (DC) systems address the storage challenge. Traditionally, these functions have been treated as separate systems with independent optimization objectives. However, modern IoT applications increasingly demand a unified approach where both MEC and DC are jointly optimized to maximize the resource utilization and minimize the operation of redundant facilities. For example, in the scenarios such as uncrewed mining \cite{park2020}, remote environmental monitoring \cite{coulby2021}, and post-disaster rescue \cite{sun2024a}, devices face dual demands. Specifically, they are required to perform computationally intensive tasks such as image recognition, anomaly detection, or path planning. Simultaneously, they are required to continuously transmit sensor or vision data for remote monitoring and model updating. While the existing terrestrial infrastructures (\textit{e.g.}, Wi-Fi, edge servers) can provide high-throughput connectivity and low-latency computing for urban deployments, their reliance on the fixed infrastructures creates significant coverage gaps in remote regions and renders them vulnerable during natural disasters and emergency scenarios where rapid deployment and resilient connectivity are paramount \cite{Li2024a}.

\par Although satellite and autonomous aerial vehicle (AAV)-assisted networks have been considered to assist MEC and DC systems, their individual deployments exhibit inherent limitations in the scenarios that require both MEC and DC. Specifically, satellite platforms, particularly the low Earth orbit (LEO) satellite constellations, can provide seamless global connectivity, which is crucial for offering stable connections and computing services in remote or disaster-stricken areas, which has made them popular in recent years \cite{song2021,jia2021a,Li2024a}. However, their orbital altitude introduces inherent latency constraints that can significantly impact the time-sensitive MEC applications requiring immediate response. Conversely, AAVs have drawn extensive attention due to their flexibility and low cost \cite{huang2025, chen2023, Li2024b}. Moreover, AAVs can hover over GDs to provide more reliable and real-time services than satellites. Nevertheless, they are severely constrained by the limited energy and onboard computing resources. These inherent shortcomings of individual systems become crucial in environments that demand intensive computation offloading and continuous DC. Fortunately, the integration of aerial and space access networks can complement each other through the collaboration of AAVs and satellites, providing on-demand communications, adaptive DC, and dynamic MEC services, which is an important application of the sixth-generation (6G) network \cite{Jia2021}.

\par However, implementing such satellite-AAV-enabled joint MEC-DC systems face several major challenges. First, the functional heterogeneity between MEC and DC introduces conflicting requirements. Specifically, MEC tasks are latency-sensitive, computation-intensive, and priority-driven, whereas DC tasks are latency-insensitive, demand high throughput, and are cost-sensitive \cite{Yang2025}. This heterogeneity complicates the allocation of communication, computation, and energy resources under dynamic network conditions. Second, the dual objectives of MEC and DC inherently compete in trajectory planning, resource allocation, and energy management. For instance, moving AAVs toward latency-critical MEC areas ensures timely task offloading but increases interference and reduces the data rate for DC devices. Processing MEC tasks locally can reduce delay. However, it consumes additional energy that could be available for propulsion or DC transmission. Furthermore, limited AAV communication and computing resources introduce trade-offs between task latency and coverage. These interconnected challenges demand innovative solutions that can handle high-dimensional decision spaces and adapt to rapidly changing operational environments.

\par Accordingly, this paper investigates the joint optimization problem of MEC and DC in satellite-AAV networks, aiming to enhance the performance of both MEC and DC while considering the energy consumption of AAVs. However, optimizing such a heterogeneous dynamic system requires online decision-making capabilities that can adapt to varying network conditions and service demands in real-time. Furthermore, the high-dimensional hybrid decision space introduces additional complexity that conventional online algorithms cannot effectively address. Therefore, this paper proposes a generative diffusion model-enabled reinforcement learning-based approach to tackle these optimization challenges by leveraging its ability to learn complex patterns and make rapid decisions. The main contributions of this paper are summarized as follows:

\begin{itemize}
    \item \textit{Satellite-AAV-Enabled Joint MEC and DC System:} We investigate a satellite-AAV-assisted system for joint MEC and DC to coordinate satellites and AAVs in performing MEC and DC simultaneously. Specifically, the system integrates an LEO satellite and AAVs to balance the computational load while enhancing the capability of the system for processing computationally intensive tasks and freshness-insensitive data for application in IoT scenarios such as factory monitoring and smart grids. To the best of our knowledge, this joint MEC and DC system in satellite-AAV-assisted networks has not yet been studied in the literature.

    \item \textit{Energy-sensitive MEC-DC Joint Optimization Problem Formulation:} We model the system to explore its heterogeneous resource allocation and dynamics, and find that the end-to-end (E2E) delay of MEC, the amount of collected data for DC, and the energy consumption of AAVs are three conflicting objectives. Therefore, we formulate a joint optimization problem to optimize these objectives simultaneously. Specifically, by adjusting the AAV movement, GD association, and offloading decisions as well as the bandwidth allocation of AAVs, we aim to minimize the average E2E delay and AAV energy consumption while maximizing the amount of collected data. Furthermore, the formulated optimization problem is a non-convex mixed-integer nonlinear programming (MINLP) problem with long-term optimization objectives and dynamics.

    \item \textit{Diffusion-Based Deep Reinforcement Learning (DRL) Solution:} To address the challenging optimization problem, we reformulate it as a Markov decision process (MDP) with a transformed action space and solve it by using the DRL framework. Specifically, we propose a diffusion-based DRL solution. The solution integrates a Gale-Sharpley (GS)-based GD association strategy into the DRL framework based on the Q-weighted variational policy optimization (QVPO) algorithm. Moreover, we employ a transformation method to convert the discrete and varying dimensional decision variables into suitable actions for DRL. The proposed QVPO-based joint AAV movement control, GD association, offloading decision, and bandwidth allocation (QAGOB) approach provides a systematic framework for the first time for the satellite-AAV joint MEC-DC scenario and realizes an effective solution.

    \item \textit{Performance Evaluation and Analysis:} Simulation results demonstrate the effectiveness and robustness of the proposed diffusion-based DRL solution. Specifically, the proposed method outperforms the other five benchmark algorithms in terms of the three optimization objectives as well as the completion rate of MEC and DC. Moreover, the QAGOB approach exhibits training stability and convergence robustness under various random environments and parameter settings. Furthermore, QAGOB improves the AAV energy consumption while achieving improvements of 11.48\% in the MEC delay and 13.99\% in the collected data volume, respectively.
\end{itemize}

\par The rest of this paper is structured as follows. Section~\ref{sec:related works} reviews the related research. Section \ref{sec:system_model} presented the detailed system models. The formulation of the joint optimization problem is presented in Section \ref{sec:problem formulation}. Section \ref{sec:proposed algorithm} proposes the diffusion-based DRL approach, and Section \ref{sec:experiments_and_analysis} provides the simulation results and analysis. Finally, the conclusion of this paper is given in Section~\ref{sec:conclusion}.

%
%%%%%%%%%%%%%%%%%%%%%% Related works %%%%%%%%%%%%%%%%%%%%%%%%%%%%%%%%%
%
\begin{table*}[htbp]
\scriptsize
\caption{Differences between This Work and Existing Works}
\label{tab: comparison}
\begin{tabular*}{\textwidth}{@{}@{\extracolsep{\fill}}ccccccccc@{}} \toprule
                                          \textbf{References}  & \textbf{AAV} & \textbf{Satellite-AAV} & \textbf{MEC} & \textbf{DC}  & \textbf{Joint MEC-DC} & \textbf{\begin{tabular}[c]{@{}c@{}}Energy \\ sensitive\end{tabular}} & \textbf{DRL} & \textbf{\begin{tabular}[c]{@{}c@{}}Diffusion-based \\ DRL\end{tabular}} \\ \midrule
\cite{Jia2021}, \cite{Gu2020}                & $\checkmark$ & $\checkmark$           & $\times$     & $\checkmark$ & $\times$              & $\checkmark$              & $\times$     & $\times$                     \\
\cite{Tun2025}, \cite{Lin2023}, \cite{Chao2020} & $\checkmark$ & $\checkmark$           & $\checkmark$ & $\times$     & $\times$              & $\checkmark$              & $\times$     & $\times$                     \\
\cite{Lei2025}, \cite{Pervez2025}, \cite{Luglio2022} & $\checkmark$ & $\checkmark$           & $\checkmark$ & $\times$     & $\times$              & $\times$                  & $\times$     & $\times$                     \\
\cite{zhang2025a}                               & $\checkmark$ & $\checkmark$           & $\times$     & $\checkmark$ & $\times$              & $\checkmark$              & $\checkmark$ & $\times$                     \\
\cite{zhang2024}                              & $\checkmark$ & $\checkmark$           & $\times$     & $\checkmark$ & $\times$              & $\times$                  & $\checkmark$ & $\times$                     \\
\cite{Gao2024a}                              & $\checkmark$ & $\checkmark$           & $\checkmark$ & $\checkmark$ & $\times$              & $\checkmark$              & $\times$     & $\times$                     \\
\cite{Liu2024}, \cite{Lakew2024}              & $\checkmark$ & $\checkmark$           & $\checkmark$ & $\checkmark$ & $\times$              & $\checkmark$              & $\checkmark$ & $\times$                     \\
\cite{Lu2024}                              & $\checkmark$ & $\checkmark$           & $\checkmark$ & $\times$     & $\times$              & $\checkmark$              & $\times$     & $\times$                     \\
\cite{Yin2024}                              & $\checkmark$ & $\checkmark$           & $\times$     & $\checkmark$ & $\times$              & $\times$                  & $\times$     & $\times$                     \\
\cite{Wang2023}, \cite{Chen2021}              & $\checkmark$ & $\checkmark$           & $\times$     & $\checkmark$ & $\times$              & $\checkmark$              & $\times$     & $\times$                     \\
\cite{Hu2023}                              & $\checkmark$ & $\checkmark$           & $\times$     & $\times$     & $\times$              & $\checkmark$              & $\times$     & $\times$                     \\
\cite{Chai2023}, \cite{Zhou2021}              & $\checkmark$ & $\checkmark$           & $\checkmark$ & $\times$     & $\times$              & $\checkmark$              & $\checkmark$ & $\times$                     \\
\cite{Lyu2023}                              & $\times$     & $\times$               & $\checkmark$ & $\times$     & $\times$              & $\checkmark$              & $\checkmark$ & $\times$                     \\
\cite{zhang2023}                              & $\checkmark$ & $\checkmark$           & $\checkmark$ & $\times$     & $\times$              & $\checkmark$              & $\checkmark$ & $\times$                     \\
\cite{guo2023}                              & $\checkmark$ & $\checkmark$           & $\times$     & $\times$     & $\times$              & $\times$                  & $\checkmark$ & $\times$                     \\
\cite{Liang2025}                              & $\checkmark$ & $\times$               & $\times$     & $\checkmark$ & $\times$              & $\checkmark$              & $\checkmark$ & $\checkmark$                 \\
\cite{zhang2025}                              & $\checkmark$ & $\times$               & $\times$     & $\times$     & $\times$              & $\checkmark$              & $\checkmark$ & $\checkmark$                 \\
\cite{Du2024}                              & $\checkmark$ & $\times$               & $\times$     & $\times$     & $\times$              & $\times$                  & $\checkmark$ & $\checkmark$                 \\
\textbf{This work}                          & $\checkmark$ & $\checkmark$           & $\checkmark$ & $\checkmark$ & $\checkmark$          & $\checkmark$              & $\checkmark$ & $\checkmark$                 \\ \bottomrule                            
\end{tabular*}
\end{table*}

\section{Related Work}
\label{sec:related works}

\par In this section, we review the related works on the satellite-AAV joint MEC-DC architecture, joint optimization for MEC, DC, and AAV energy consumption, as well as the optimization methods of the DRL framework. Moreover, Table~\ref{tab: comparison} illustrates the differences between the state-of-the-art works and this work.

\subsection{Satellite-AAV-Enabled MEC and DC Architecture}

\par Recent studies have explored satellite-AAV collaborative architectures for MEC to address the limitations of terrestrial infrastructure in remote areas. For instance, the authors in \cite{Tun2025} investigated an integrated space-air-ground network leveraging the terahertz (THz) spectrum and AAV collaboration, where AAVs act as reconfigurable MEC nodes to optimize task offloading and resource allocation for energy-efficient service delivery. The authors in \cite{Lei2025} introduced an edge information hub architecture, deploying AAVs as aerial coordinators to orchestrate satellite-backhauled sensing data and MEC resources for the control of ground robots in disaster scenarios. The authors in \cite{Pervez2025} investigated a multi-AAV satellite-aerial-terrestrial network, where AAVs and satellites jointly serve ground users through task splitting, trajectory planning, and computation resource distribution. The authors in \cite{Lin2023} designed a tactical ad-hoc MEC network combining LEO satellites, AAV-MEC servers, and ground nodes to enable resilient computation offloading in hostile environments. For IoT applications in isolated regions, the authors in \cite{Luglio2022} demonstrated a satellite-based MEC architecture with AAV-assisted edge preprocessing to reduce backhaul traffic and latency for delay-tolerant tasks. The authors in \cite{Chao2020} further extended this paradigm to vehicular networks, where AAVs dynamically offload tasks to satellites or terrestrial MEC servers to alleviate congestion in mobile scenarios.

\par Prior works have also investigated satellite-AAV-assisted DC systems. For example, the authors in \cite{Jia2021} proposed an LEO satellite-assisted AAV framework for the Internet of remote Things (IoRT) DC, where AAVs gather data from ground sensors and employ dual transmission modes to optimize the energy consumption and trajectory design. The authors in \cite{Gu2020} integrated fault-tolerant coding into a satellite-AAV mobile edge caching system, where LEO satellites broadcast data to AAVs for the decentralized sensor DC. The authors in \cite{zhang2025a} addressed information freshness in space-air-ground integrated networks (SAGINs) by designing a multi-layer architecture combining satellites, high-altitude platforms (HAPs), and AAVs. The authors in \cite{zhang2024} introduced a cache-enabled AAV-LEO satellite system to mitigate backhaul link fluctuations, utilizing AAV-mounted cache nodes for temporary data storage and DRL-based trajectory optimization to adapt to the uneven data distributions in dynamic environments. The authors in \cite{Gao2024a} tackled massive data aggregation challenges in multi-tier satellite-AAV networks by proposing a game-theoretic multiple access approach.

\par Some studies have involved both MEC and DC. For example, the authors in \cite{Liu2024} proposed a multi-AAV-assisted air-ground-space power IoT (PIoT) framework that deeply integrates wireless power transfer (WPT), MEC, and DC. In this framework, AAVs provide energy to ground power devices through WPT and collect data, then process the data in real time using onboard MEC resources, and finally transmit the processed data via LEO satellites.

\par In summary, existing research on satellite-AAV-assisted networks has primarily focused on independent MEC and DC systems. With satellites and AAVs, jointly optimizing MEC and DC introduces additional challenges, such as resource competition, conflicts among multiple objectives, and scheduling issues under dynamic heterogeneous channel conditions. As such, this motivates us to explore joint MEC and DC in satellite-AAV-enabled networks.

\subsection{Joint Optimization for MEC and DC}

\par For the optimization of satellite and AAV-enabled networks, existing research has primarily focused on resource allocation and AAV trajectory optimization. For example, the authors in \cite{Lu2024} focused on minimizing the total energy consumption of GDs in an AAV-assisted satellite MEC system, where the joint optimization of the task offloading decisions and AAV trajectory was performed under energy budget constraints. The authors in \cite{Yin2024} prioritized secure uplink communications in satellite-supported IoT networks, formulating a max-min secrecy rate optimization problem that jointly optimizes the uplink power allocation, AAV beamforming, and positioning to mitigate eavesdropping risks. The authors in \cite{Wang2023} aimed to maximize energy efficiency in non-orthogonal multiple access (NOMA)-enabled SAGINs by co-designing the user association, power allocation, and three-dimensional (3D) AAV trajectory while accounting for the satellite mobility. The authors in \cite{Chen2021} addressed energy-constrained computation offloading in SAGINs, minimizing the system latency under task arrival uncertainty by optimizing the task offloading proportions and AAV-to-server forwarding decisions. 

\par Some other works have investigated the joint optimization of multiple objectives. For instance, the authors in \cite{Hu2023} investigated an AAV-enabled SAGIN where AAVs act as base stations or MEC servers, jointly optimizing the 3D trajectory and resource allocation to balance the user quality-of-experience (QoE) and energy efficiency. The authors in \cite{Chai2023} studied the joint computation offloading problem in the satellite-AAV framework, minimizing the task delay and the task energy consumption while reducing the total task costs.

\par However, the existing optimization approaches in these studies do not jointly optimize the MEC and DC objectives or consider the trade-offs between the MEC latency, the amount of collected data, and the AAV energy consumption. Therefore, they are not suitable for our considered system.

\subsection{Optimization Approaches for MEC and DC}

\par To overcome the challenges of resource allocation and dynamic scheduling in complex network optimization, DRL methods are considered a promising solution and have been increasingly applied to satellite-AAV-assisted communication systems. For example, the authors in \cite{Lakew2024} proposed a mixed discrete-continuous DRL algorithm for task offloading in LEO-MEC-assisted energy-harvesting AAV systems, enabling adaptive decision-making under the time-varying satellite connectivity and energy constraints. The authors in \cite{Zhou2021} developed a risk-sensitive DRL framework for delay-oriented IoT task scheduling in SAGINs, balancing energy consumption and latency by evaluating state-specific risks in energy-constrained MDPs. The authors in \cite{Lyu2023} investigated multi-agent collaborative LEO-IoT scenarios with a partially observable Markov decision process (POMDP) model and proposed information broadcasting mechanisms to optimize the task offloading latency and energy efficiency across distributed satellites and AAVs. The authors in \cite{zhang2023} further integrated multi-agent actor-critic frameworks to coordinate orbital edge offloading strategies among AAVs and LEO satellites, maximizing the task completion rates under delay constraints while minimizing the AAV energy consumption. The authors in \cite{Wang2025a} proposed a DRL-based multi-objective evolutionary algorithm to adjust the selection of evolutionary operators according to the dynamics of the optimization objectives. The authors in \cite{guo2023} applied multi-agent deep deterministic policy gradient (MADDPG) in cognitive satellite-AAV networks, jointly optimizing the AAV trajectory and power allocation for spectrum-efficient NOMA transmissions under strict delay-sensitive QoS requirements.

\par While traditional DRL methods are capable of solving optimization problems in real-time decision-making scenarios, their policy networks struggle to capture patterns and dependencies in complex data, limiting the policy expression and exploration capabilities \cite{Liang2025}. In contrast, diffusion-based DRL methods have recently received increasing attention. Compared with diffusion models in image generation, their application objectives shift from generative data synthesis to policy optimization, their outputs transition from pixel representations to action vectors, and their training paradigm evolves from unsupervised learning to value-guided reinforcement learning with explicit reward optimization \cite{Du2024}. For example, the authors in \cite{Liang2025} proposed a diffusion model-enhanced DRL framework for the AAV-assisted secure DC and energy transfer in IoT, where a diffusion model generates the AAV trajectories and device scheduling strategies under jamming attacks, balancing the secure age of information (AoI) minimization and energy efficiency. The authors in \cite{zhang2025} investigated AAV swarm-enabled secure surveillance systems, integrating the generative diffusion models into DRL to optimize collaborative beamforming and trajectory planning for the joint secrecy rate maximization and flight energy reduction. By leveraging the probabilistic reasoning capabilities of diffusion models, their approach captures the complex trade-offs between multi-objective constraints in the dynamic aerial environments. The authors in \cite{Du2024} further demonstrated the versatility of diffusion-based DRL in network traffic management, where diffusion models guide the adaptive service provider selection for personalized content generation, showcasing potential extensions of diffusion models. 

\par Although these diffusion-based DRL methods have shown potential in real-time decision-making scenarios, they do not consider the low sample efficiency and limited exploration capabilities of diffusion models when they are directly applied to online DRL methods. Moreover, the aforementioned works are all purely discrete or continuous, which cannot be applied to the hybrid and variable-dimensional action space of the formulated joint optimization in satellite-AAV-assisted communication systems. To address these limitations, this paper seeks to propose an efficient diffusion-based DRL solution to accommodate the hybrid and variable-dimensional action spaces, thus meeting the specific requirements of satellite-AAV-enabled MEC and DC optimization problems.

%
%System model and problem formulation
%
\section{System Model}
\label{sec:system_model}

\par In this section, we present the considered satellite-AAV-enabled joint MEC-DC network, including the network model, communication model, MEC model, and DC model.

\begin{figure}[tbp]
\centering
\includegraphics[width=0.49\textwidth]{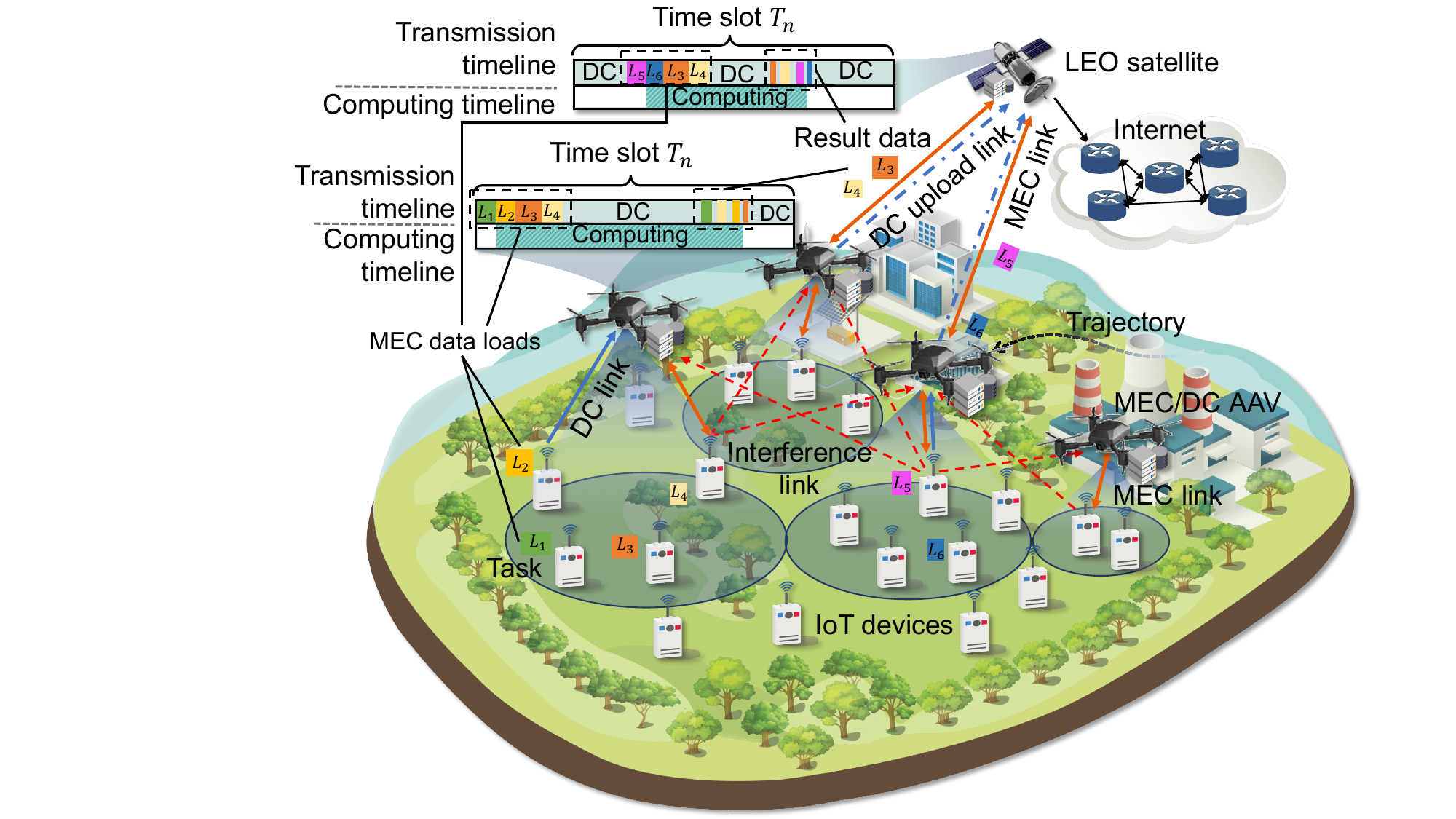}
\caption{Illustration of the LEO satellite-AAV-enabled joint MEC-DC system. The system enables multi-tier computing where tasks are processed by AAVs or further offloaded to the satellite. The satellite also acts as the destination for data collection. Note that inter-cell interference exists among AAVs during the simultaneous service provisioning.}
\label{fig:network_model}
\end{figure}

\subsection{Network Model}

\par Fig. \ref{fig:network_model} illustrates the considered satellite-AAV-enabled joint MEC-DC system. In this system, there are $N_G$ stationary IoT GDs in a monitored area, denoted as $g \in \mathcal{G} = \{1, 2, \dots, N_G\}$, and these GDs have different functions and require performing computing or uploading data. For example, the GDs may need to perform computation-intensive tasks with tight time constraints, such as fire warning \cite{ding2024a} and image processing \cite{ding2023}. Moreover, these GDs continuously store data during operation and need to be uploaded or reported for further analysis. However, since the GDs in the considered scenario are in remote areas, there is no communication infrastructure that can directly provide communication and computing services. Moreover, the GDs have limited computing capability and restricted storage capacity, hence, the data needs to be transmitted to a nearby edge server or data center for processing or storage. Therefore, there is a set of AAVs, denoted as $v \in\mathcal{V} = \{ 1, 2, \dots,N_V\}$, equipped with computing units and storage units to provide edge computing and data collecting service for the GDs. Following the edge computing and data collection (DC), the result data is returned to the GDs, and the collected data is uploaded to an LEO satellite, denoted as $S_a$, which can provide backhaul connectivity to the Internet and has computing capability to complement the MEC.

\par In a system operation cycle, the time horizon is divided into $T$ slots with a length of $\tau$, which can be denoted as $t \in \mathcal{T}=\{1, 2, \dots, T\}$. In each time slot $t$, the GDs accumulate the data waiting to be collected and generate computing tasks at irregular intervals. Note that the AAVs are equipped with an omnidirectional antenna and a directional antenna to communicate with the GDs and the satellite, respectively. Subsequently, each AAV flies or hovers over the GDs to provide edge computing service and collect stored data. Specifically, as shown in Fig. \ref{fig:network_model}, the AAVs first transmit the MEC task data and start collecting the stored data of the nearby GDs while they are performing computing, and the transmission of MEC task data is finished. Note that we consider that the AAVs do not have the full-duplex function \cite{guo2021}, and thus, the transmission of DC will be interrupted when MEC result data requires returning. At the same time, the AAVs can decide whether to offload part of the MEC tasks to the satellite $S_a$ to alleviate the computational workload. Since GDs usually have limited transmit power and cannot communicate with the satellite directly, the result data will be returned to the AAV and further back to the GD after the satellite computation is completed. 

\par Moreover, since the AAVs communicate with the GDs under the same frequency band, there exists co-channel interference that needs to be considered. Furthermore, since IoT devices are generally difficult to apply one-to-many communication protocols due to their limited transmit power and computation resources, we consider that a GD can only be served by one AAV while an AAV can serve multiple GDs within a time slot~\cite{chen2023}. Additionally, the orthogonal frequency division multiple access (OFDMA) protocol is applied in each AAV to avoid inner-cell interference \cite{zheng2022}.

\par Without loss of generality, we consider a 3D Cartesian coordinate system, where the AAVs move at a fixed altitude $H$. Therefore, the positions of AAVs and GDs in time slot $t$ are denoted as $P_v(t) = \{x_v(t),y_v(t),H\}$ and $P_g(t)=\{x_g(t),y_g(t),0\}$, respectively.

\subsection{Communication Model}

\par Based on the aforementioned system, the considered air-ground channel model and the satellite-air channel model are introduced. 

\subsubsection{Ground-to-Air (G2A) / Air-to-Ground (A2G) Channel Model}

\par In each time slot, the channels between the GDs and AAVs are considered quasi-static and vary across different time slots~\cite{li2025}. Subsequently, the AAVs can establish line-of-sight (LoS) links with the GDs due to their high flight altitude \cite{sun2024a}. Therefore, the probabilistic LoS model is adopted in this work and the probability of Los link is given by $P_{v,g}^{LoS}(t) = 1/(1+n_1 \exp\{-n_2((180/\pi)\arctan(H/d_{v,g}(t)) - n_1)\})$, where the distance between AAV $v$ and GD $g$ is represented by $d_{v,g}(t)=\left \| P_v(t) - P_g(t) \right \|$, $n_1$ and $n_2$ are constant values related to the environment~\cite{guo2023}. Accordingly, the path loss between AAV $v$ and GD $g$ is given by
\begin{equation}
    PL_{v,g}(t) = L_{v,g}(t) + P_{v,g}^{LoS}(t)\delta_{LoS} + P_{v,g}^{NLoS}(t)\delta_{NLoS},
\end{equation}
where $L_{v,g}(t) = 20\log d_{v,g}(t)+20\log f_c+20\log(\frac{4\pi}{c})$ is the free space path loss between AAV $v$ and GD $g$ in time slot~$t$ and $f_c$, as well as $c$, are the carrier frequency and velocity of light, respectively. Moreover, $P_{v,g}^{NLoS}(t)=1-P_{v,g}^{Los}(t)$ is the probability of a non-LoS (NLoS) link between AAV $v$ and GD $g$ in time slot $t$, $\delta_{LoS}$ and $\delta_{NLoS}$ are the constants corresponding to the excessive path losses for LoS and NLoS links, respectively.

\par As a result, the channel gain between AAV $v$ and GD $g$ is given by
\begin{equation}\label{eq:gain}
\begin{aligned}
h_{v,g}(t) &= 10^{-\frac{PL_{v,g}(t)}{10}}\\
           &= \frac{10^{-\frac{(\delta_{LoS}-\delta_{NLoS})P_{v,g}^{LoS}(t)+\delta_{NLoS}}{10}}}{\left (\left \|P_v(t)-P_g(t)\right \| \right )^2\left (\frac{4\pi f_c}{c}\right )^2}.
\end{aligned}    
\end{equation}

\par To represent the association state between the AAVs and GDs, we define an indicator variable $X_{v,g}(t) = 1$ or $0$ to denote whether AAV $v$ is associated with GD $g$. As such, the transmission rate from GD $g$ to AAV $v$ is given by $R_{v,g}^{G2A}(t) = B_{v,g}(t)\log_2\left (1+\frac{p_g(t)h_{v,g}(t)}{I_m+n_0B_{v,g}(t)}\right )$, where $p_g(t)$ is the transmit power of GD $g$ in time slot~$t$, $I_m={\textstyle \sum_{j=1,j\ne v}^{M}}{\textstyle \sum_{l=1,l\ne g}^{N}}X_{j,l}(t)p_{j,l}(t)h_{j,l}(t)$ is the co-channel interference from other GDs associated with other AAVs in time slot $t$, $n_0$ is the noise power spectral density, and $B_{v,g}(t)$ denotes the bandwidth that AAV $v$ allocates to GD $g$ in time slot $t$.

\par Since the result data (\textit{e.g.}, face recognition and image processing) is generally much less than the MEC task data \cite{chen2023}, AAV downlink transmission durations are very short. Therefore, the inter-cell interference from other AAVs can be ignored for simplicity, allowing consideration of interference-free downlink transmission \cite{Liao2025}. As such, the transmission rate from AAV $v$ to GD $g$ is given by $R_{v,g}^{A2G}(t) = B_{v,g}(t)\log_2\left (1+\frac{p_v(t)h_{v,g}(t)}{n_0B_{v,g}(t)}\right )$, where $p_v(t)$ is the transmit power of AAV $v$ in time slot $t$.

\subsubsection{Air-to-Satellite (A2S) / Satellite-to-Air (S2A) Channel Model}

\par In the process of communication between the satellite and AAVs, we consider the impact of rain attenuation, which can be formulated based on the Weibull distribution \cite{Kanellopoulos2014}. Therefore, the power attenuation of the satellite-AAV link is given by
\begin{equation}
    \gamma_{v,s}(t)=\frac{\lambda^2G_vG_s}{(4\pi d_{v,s}(t))^2}10^{-\frac{F_{rain}}{10}},
\end{equation}
where $\lambda$ is the carrier wave length, $G_v$ and $G_s$ are the antenna gains of AAV $v$ and satellite $S_a$, respectively. Moreover, $d_{v,g}(t)$ represents the distance between AAV $v$ and satellite $S_a$, and $F_{rain}$ is the rain attenuation modeled by Weibull distribution \cite{Kanellopoulos2014}. Thus, the transmission rate from AAV $v$ to satellite $S_a$ is given by $R_{v,s}^{A2S}(t)=B_{v,s}(t)\log_2 \left ( 1 + \frac{p_v(t)\gamma_{v,s}(t)}{n_0B_{v,s}(t)}   \right )$, where $B_{v,s}(t)$ is the bandwidth that satellite $S_a$ allocates to AAV $v$ in time slot $t$, and its value is equal to the total satellite bandwidth $B_s$ divided by the number of AAVs connected to satellite $S_a$ in time slot $t$.

\par Similarly, the transmission rate from satellite $S_a$ to AAV $v$ is given by $R_{v,s}^{S2A}(t)=B_{v,s}(t)\log_2 \left ( 1 + \frac{p_s(t)\gamma_{v,s}(t)}{n_0B_{v,s}(t)}   \right )$, where $p_s(t)$ is the transmit power of satellite $S_a$ in time slot~$t$.

\subsection{MEC Model}

\par The MEC tasks generated by the GDs have their inherent properties in the considered system. Specifically, the MEC tasks generally have a processing time constraint, which is set to ensure the quality of service (QoS) of the GDs. Moreover, in the considered scenario, the communication and computation resources are limited, thus, the MEC tasks may not be processed immediately but have a deadline, such as \cite{zhan2021}. Therefore, a task is considered valid when it starts to be processed within the deadline, otherwise, the task is invalid. Based on this, we denote the set of tasks of GD $g$ as $f_{g,k} \in \mathcal{F}=\{f_{g,1},f_{g,2},\dots,f_{g,K_g}\}$, where $K_g$ denotes the number of tasks of GD $g$ and a four-element tuple $\{L_{g,k},T_{g,k}^{max},D_{g,k}, \rho_{g,k}\}$ is used to characterize the MEC task $f_{g,k}$, where $L_{g,k}$, $T_{g,k}^{max}$, and $D_{g,k}$ represent the data length, the maximum tolerable delay, and the deadline of the task $f_{g,k}$, respectively. Moreover, $\rho_{g,k}$ is the ratio of the result data size to the task data size, which is in the value interval of $(0,1)$.

\subsubsection{Task Generation}

\par In real-world scenarios, the MEC tasks usually randomly arrive and are unpredictable. To simulate this randomness, existing studies typically model task generation by sampling from a Bernoulli distribution or by assuming a normally distributed generation probability for each time slot \cite{wei2021,li2024}. However, these static assumptions often result in scenarios that are either consistently dense or sparse. Such lack of workload volatility limits the ability of algorithms to learn robust strategies, potentially requiring complex manual reward shaping to accommodate extreme cases. Additionally, these approaches assume that the task generation probability is constant and independent across time slots, which fails to capture the dynamic nature of real-world task arrivals. Therefore, to better reflect the stochastic nature of task arrivals, we adopt the exponential distribution \cite{Han2023}, which is widely used to model the random inter-arrival times of computation tasks in MEC systems. Consequently, the task generation model can be given by
\begin{equation}
    P_{g,f}(t) = 1-e^{-\delta_f t},
\end{equation}
where $P_{g,f}(t)$ is the probability that GD $g$ generates an MEC task $f$ in time slot $t$, $\delta_f$ is the task density coefficient, which controls the speed of task generation. Moreover, to simulate the stochastic nature of the length of the task data, we consider a Poisson distribution to model the size of the task data, \textit{i.e.}, $L_{g,k} \sim P(\lambda_{MEC})$.

\subsubsection{End-to-End (E2E) Delay}

\par The MEC delay mainly consists of three parts, which are the uplink transmission delay, the computation delay, and the downlink transmission delay, respectively. 

\par At the beginning of the MEC process, the task data is first uploaded to the AAVs from the GDs and the uplink transmission delay from GD $g$ to AAV $v$ is given by $T_{v,g,k}^{G2A}(t)= X_{v,g}(t)L_{g,k}/R_{v,g}^{G2A}$. Subsequently, if AAV $v$ has requirement that task $f_{g,k}$ needs to be further offloaded to satellite $S_a$, the uplink transmission delay from AAV $v$ to satellite $S_a$ is given by $T_{s,v,g,k}^{A2S}(t) = L_{g,k}/R_{v,s}^{A2S}(t)$.

\par After the transmission of task data, the MEC tasks are computed on the AAVs or the satellite. Moreover, we consider that the GDs have limited computational resources, and thus the local computing is unavailable and the tasks are considered to be offloaded only to the AAVs or the satellite \cite{yu2022}. Therefore, when task $f_{g,k}$ is offloaded to AAV $v$ to be computed, the computation delay in time slot $t$ is given by $T_{v,g,k}^{com}(t) = X_{v,g} I_vL_{g,k}/\omega_v$, where $I_v$ is the computation intensity on AAV $v$ (cycles/bit) and $\omega_v$ denotes the computing capability, \textit{i.e.}, the CPU operating frequency of AAV $v$. Similarly, the computation delay on satellite $S_a$ in time slot $t$ is given by $T_{s,g,k}^{com}(t) = I_s L_{g,k}/\omega_s$, where $I_s$ and $\omega_s$ are computation intensity on satellite $S_a$ and computing capability of satellite $S_a$, respectively.

\par Finally, the result data is transmitted back to the GDs from the satellite or the AAVs. Specifically, the downlink transmission delay from satellite $S_a$ to AAV $v$ is given by $T_{s,v,g,k}^{S2A}(t) = \rho_{g,k}L_{g,k}/R_{v,s}^{S2A}(t)$. Subsequently, the downlink transmission delay from AAV $v$ to GD $g$ is given by $T_{v,g,k}^{A2G}(t)= X_{v,g}\rho_{g,k}L_{g,k}/R_{v,g}^{A2G}(t)$.

\par Therefore, in the case that task $k$ only be offloaded to AAV $v$ to be processed, the delay for task $k$ in time slot $t$ can be given by
\begin{equation}
    T_{v,g,k}(t) = T^{G2A}_{v,g,k}(t) + T^{com}_{v,g,k}(t)+T^{A2G}_{v,g,k}(t).
\end{equation}

\par Moreover, in the case that task $k$ is further offloaded to AAV $v$ to be processed, the delay for task $k$ in time slot $t$ can be given by
\begin{equation}
\begin{aligned}
    T_{s,v,g,k}(t) = &T^{G2A}_{v,g,k}(t) + T^{A2S}_{s,v,g,k}(t) + T^{com}_{s,g,k}(t) + T^{S2A}_{s,v,g,k}(t) +\\ &T^{A2G}_{v,g,k}(t) + 2T^{pro}_{s,v}(t),
\end{aligned}
\end{equation}
where $T^{pro}_{s,v}(t) = d_{v,s}(t)/c$ is the propagation delay.

\par As a result, the total E2E delay of tasks offloaded from the GDs to the AAV $v$ in time slot $t$ is given by

\begin{equation}
    T_{v}(t) = \sum\nolimits_{g=1}^{N_G}(1-o_g(t))T_{v,g,k}(t)+\sum\nolimits_{g=1}^{N_G}o_g(t)T_{s,v,g,k}(t),
\end{equation}
where $o_g(t)$ is a offloading variable, which denotes whether the task $f$ from GD $g$ is further offloaded to the satellite.

\subsubsection{MEC Task completion rate}

\par In IoT MEC scenarios, tasks are typically latency-sensitive and subject to strict delay constraints. A task is considered failed if it cannot be completed within its maximum tolerable latency due to the weak communication signal, heavy computational loads, or insufficient CPU computing capability on the nearby edge servers. Therefore, the task completion status is an important metric to reflect the service coverage capability and the QoS of the GDs. Denote $b_{g,k}$ as the completion status indicator variable with a value of 1 or 0 to indicate whether the task $f_{g,k}$ is completed or not, respectively. As such, the task completion rate in the system is defined as follows:
\begin{equation}
    \mathcal{C}^{MEC} = \frac{\sum\nolimits_{g=1}^{N_G}\sum\nolimits_{k=1}^{K_g}b_{g,k}}{\sum_{g=1}^{N_G}K_g} \times 100 \%.
\end{equation}

\subsection{DC Model}

\par Recall that the GDs accumulate data in every time slot, and the AAVs need to collect the delay-insensitive data and upload it to the satellite to return to the cloud data center. Therefore, the local data accumulation of the GDs is considered to be generated continuously in each time slot, and the amount of data generated by GD $g$ in time slot $t$ follows a Poisson distribution, namely $D_g(t) \sim P(\lambda_{DC})$ \cite{Lee2024}.

\par Subsequently, the accumulated data is collected and temporarily stored by the AAVs before being uploaded to the satellite. Notably, the collected data is delay-insensitive, which allows AAVs to defer immediate upload. Furthermore, given the potential significant distance from cloud data centers, the AAVs first handle the transmission of the MEC task data, and only collect and upload data to the satellite when the MEC tasks are not being transmitted. Therefore, the amount of collected data is defined as the amount of data received by satellite $S_a$, which is given by
\begin{equation}D(t)=\sum\nolimits_{v=1}^{N_V}\tau^{DC}_v(t)R^{A2S}_{v,s}(t),
\end{equation}
where $\tau^{DC}_v(t)$ denotes the duration of AAV $v$ performing DC in time slot $t$.

\par Similar to the MEC task completion rate, we define the satellite DC rate to reflect the DC service coverage capability of the considered system. The DC rate is the ratio of the total volume of the received DC data at the satellite $S_a$ to the total amount of the data generated by all GDs, which is given by
\begin{equation}
    \mathcal{C}^{DC} = \frac{\sum\nolimits_{t=0}^TD(t)}{\sum\nolimits_{g=1}^{N_G}\sum\nolimits_{t=0}^TD_g(t)}\times100\%.
\end{equation}

\subsection{Rotary-Wing AAV Energy Consumption Model}

\par To ensure the continuous operation of the considered satellite-AAV-enabled joint MEC-DC network, the energy management of the AAVs is crucial due to the limited energy capacity of AAVs. Specifically, the energy consumption of the AAVs mainly comprises propulsion power consumption and computation energy consumption.

\par We denote $E_v^m(t)$ as the propulsion power consumption of AAV $v$ hovering and moving in a two-dimensional (2D) plane in time slot $t$, and the details of the propulsion power consumption model can refer to \cite{huang2025}. 

\par For the edge computing on AAV $v$, the energy consumed to process task $f_{g,k}$ in time slot $t$ can be given by~\cite{jiang2023}
\begin{equation}
    E_{v,g}^c(t) = \kappa_v CL_{g,k}(t),
\end{equation}
where $\kappa_v$ is the energy consumed per CPU cycle on AAV $v$.

\par Therefore, the overall energy consumption of AAV $v$ in time slot $t$ is defined as follows:
\begin{equation}
    E_{v}(t)=E_v^m(t)+\sum\nolimits_{g=1}^{N_G} E_{v,g}^c(t).
\end{equation}

%
% problem formulation
%

\section{Problem Formulation}\label{sec:problem formulation}

\par In the considered satellite-AAV joint MEC-DC system, the satellite and AAVs are considered to provide MEC and DC services for the GDs to enhance the performance of both MEC and DC under energy constraints. Therefore, the considered system involves three objectives, \textit{i.e.}, reducing the average E2E delay of the computing tasks of GDs, increasing the amount of collected data of GDs, and reducing the energy consumption of AAVs.

\par During the operation of the aforementioned system, due to the co-channel interference, the position of the AAVs, the association between the AAVs and GDs, and the bandwidth allocation of the AAVs jointly determine the communication rate between the GDs and AAVs, thereby affecting the average E2E delay and the amount of collected data. As such, these three variables are interdependent and coupled. Additionally, the offloading decision from the AAVs to the satellite can influence the computational load balancing during MEC and the computing energy consumption of the AAVs. Thus, there is an interplay and conflicting correlation among the AAV offloading decision, the AAV movement, the GD association, and the AAV bandwidth allocation. The definitions of these decision variables are as follows.

\par We define four decision variables related to the AAV: \textit{(i)} The AAV positions consisting of the continuous coordinates of each AAV are represented by the matrix $\boldsymbol{P}=\{P_v(t)| v\in\mathcal{V}, t\in \mathcal{T}\}$. \textit{(ii)} The association between the GDs and AAVs is represented by the matrix $\boldsymbol{X}=\{X_{v,g}(t)| v\in\mathcal{V}, g\in\mathcal{G}, t\in\mathcal{T}\}$ with each discrete element denotes whether the AAV-GD are associated. \textit{(iii)} The AAV offloading decision is represented by the matrix $\boldsymbol{O}=\{o_v(t)| v\in\mathcal{V}, t\in \mathcal{T}\}$ containing the vectors of the AAV service capacity size dimensions, with each discrete element of the vector representing whether to offload the task to the satellite. \textit{(iv)} The AAV bandwidth allocation is represented by the matrix $\boldsymbol{B}=\{b_v(t)| v\in\mathcal{V}, t\in \mathcal{T}\}$, where each continuous vector element represents the size of the bandwidth allocated to the GD associated with the AAV. In the following, the expressions for the considered optimization objectives are provided. Additionally, a detailed analysis of the weighted sum method employed in this paper and alternative multi-objective optimization approaches is provided in Appendix A.

\par \textit{\textbf{Optimization Objective 1:}} The first objective is to improve the performance of the AAVs and the satellite when providing the MEC services for the GDs, namely, the average E2E delay. Thus, the first objective is given by
\begin{equation}
    f_1 \left ( \boldsymbol{P}, \boldsymbol{X}, \boldsymbol{O}, \boldsymbol{B} \right ) = \frac{\sum\nolimits_{t=1}^T\sum\nolimits_{v=1}^{N_V}T_v(t)}{\sum\nolimits_{g=1}^{N_G}K_g}.
\end{equation}

\par \textit{\textbf{Optimization Objective 2:}} The second objective is to increase the amount of collected data, which is given by
\begin{equation}
    f_2 (\boldsymbol{P}, \boldsymbol{X}, \boldsymbol{O}, \boldsymbol{B}) = \sum\nolimits_{t=1}^{T}D(t).
\end{equation}

\par \textit{\textbf{Optimization Objective 3:}} The third optimization objective is to reduce the energy consumption of the AAVs during the processes of MEC and DC. Therefore, the third optimization objective is given by
\begin{equation}
    f_3 (\boldsymbol{P}, \boldsymbol{X}, \boldsymbol{O}) = \sum\nolimits_{t=1}^T\sum\nolimits_{v=1}^{N_V}E_v(t).
\end{equation}

\par According to the aforementioned three objectives, the considered optimization problem can be formulated as follows:

% \par To balance the three objectives, appropriate multi-objective optimization methods can be employed. Specifically, Pareto optimization offers a comprehensive trade-off analysis by computing a set of Pareto-optimal solutions. However, it significantly increase computational cost and training complexity. An alternative is hierarchical optimization, which establishes strict priority among objectives. This approach performs well in scenarios with rigid quality-of-service requirements, while it is less suitable for problems with tightly coupled objectives. Another option is to apply dynamic weighting, allowing flexible adjustment of objective weights during optimization. However, it increases system complexity and introduces additional hyperparameters. In this context, weighted summation can be adopted for its simplicity, interpretability, and ease of implementation, which is particularly advantageous for the diffusion-based DRL framework in this work. Therefore, the considered optimization problem can be formulated as follows:
\begin{subequations}
{\allowdisplaybreaks[4]
\begin{align}
    \mathcal{P}: \min_{\{\boldsymbol{P}, \boldsymbol{X}, \boldsymbol{O}, \boldsymbol{B}\}} &Q= \{f_1,-f_2,f_3\}, \\
    \text{s.t.}~~&T_{v,g,k}^f(t) \le T_{g,k}^{max}, \forall v \in \mathcal{V}, \forall g \in \mathcal{G},\forall t\in\mathcal{T}, \label{eq:subb}\\
    &T_{s,v,g,k}^f(t) \le T_{g,k}^{max}, \forall v \in \mathcal{V}, \forall g \in \mathcal{G},\forall t\in\mathcal{T}, \label{eq:subc}\\
    &\sum\nolimits_{v\in\mathcal{V}}X_{v,g}(t) = 1, \forall g \in \mathcal{G}, \forall t\in\mathcal{T}, \label{eq:subd}\\
    &\sum_{g\in\mathcal{G}}X_{v,g}(t) \le N_g^{max}, \forall g \in \mathcal{G}, \forall t\in\mathcal{T}, \label{eq:sube}\\
    &x_{min} \le x_v(t) \le x_{max}, \forall v \in \mathcal{V}, \forall t\in\mathcal{T}, \label{eq:subf}\\
    &y_{min} \le y_v(t) \le y_{max}, \forall v \in \mathcal{V}, \forall t\in\mathcal{T}, \label{eq:subg}\\
    &0 \le m_v(t) \le m_v^{max}, \forall v \in \mathcal{V}, \forall t\in\mathcal{T}, \label{eq:subh}\\
    &\left \| P_a(t)-P_b(t) \right \| \ge d_{min}, \forall a, b \in \mathcal{V}, a \ne b,\nonumber\\
    & \forall t\in\mathcal{T}, \label{eq:subi}
\end{align}}
\end{subequations}
where \eqref{eq:subb} and \eqref{eq:subc} impose the constraints on the delay of the tasks for the GDs, \eqref{eq:subd} and \eqref{eq:sube} constrain the ranges of the associated indicator variables and limit the maximum number of the associated AAVs. Moreover, \eqref{eq:subf} and \eqref{eq:subg} restrict the movement range of the AAVs in the X and Y directions, \eqref{eq:subh} limits the maximum movement distance of an AAV within a time slot, and \eqref{eq:subi} ensures collision avoidance among the AAVs.

\par The problem $\mathcal{P}$ exhibits the following key properties. \textit{First}, the problem $\mathcal{P}$ is an MINLP problem, containing both continuous decision variables ($\boldsymbol{P}$, $\boldsymbol{B}$) and integer variables ($\boldsymbol{X}$, $\boldsymbol{O}$), with nonlinear relationships in the constraints such as \eqref{eq:subi}. Furthermore, the problem $\mathcal{P}$ is non-convex due to its non-convex constraints including \eqref{eq:subb}, \eqref{eq:subc}, and \eqref{eq:subi} \cite{chen2023,zhang2025}. \textit{Second}, the optimization objectives of the problem $\mathcal{P}$ are influenced by the dynamic nature of the tasks from the GDs and the long-term movement state of the AAVs. This requires both immediate performance to adapt to the dynamic changes in the environment and sustainable long-term operational strategies. \textit{Finally}, the solution space of problem $\mathcal{P}$ is complex due to the high-dimensional and interdependent variables, which also exhibit multi-modal decision characteristics where the multiple distinct solutions may achieve similar performance levels.

\par Consequently, the problem $\mathcal{P}$ is a non-convex MINLP problem with long-term optimization objectives and dynamics. Although traditional methods such as convex optimization and evolutionary computing are effective in many cases, they are unsuitable for optimization in the dynamic and uncertain environments. For instance, in every time slot, traditional methods need to iterate repeatedly to learn the optimal strategy, which may lead to the excessive computational costs and be inefficient. Fortunately, DRL is capable of providing sequential dynamic decision-making and optimal control for long-term optimization problems in dynamic environments, which is a promising method \cite{zhang2025}. In the following section, the proposed solution is detailed.

\section{DRL-Based Method}
\label{sec:proposed algorithm}

\par In this section, we propose the QAGOB approach to address the considered joint optimization problem. Specifically, we first employ an MDP to reformulate the considered joint optimization problem and simplify the action space. Subsequently, the QAGOB method is proposed.

\subsection{MDP Simplification and Formulation}

\par The problem $\mathcal{P}$ can be reformulated as an MDP to facilitate the application of DRL, which typically consists of a five-tuple $<\mathcal{S}, \mathcal{A}, \mathcal{P}^{\circ}, \mathcal{R}, \gamma>$, where $\mathcal{S}$, $\mathcal{A}$, $\mathcal{P}^{\circ}$, $\mathcal{R}$, and $\gamma$ represent the state space, action space, state transition probability, reward, and discount factor, respectively.

\par Generally, the agent can be deployed in a local control center and all decision variables in the formulated optimization problem are converted into the action space in the MDP to make decisions. However, there are several main challenges in deriving the action space from the decision variables in the modeled problem $\boldsymbol{P}$. First, the decision variables include both continuous variables ($\boldsymbol{P}$, $\boldsymbol{B}$) and discrete variables ($\boldsymbol{X}$, $\boldsymbol{O}$), which are usually difficult to solve by directly using DRL algorithms. Second, although the hybrid action space can be converted into a continuous action space by mapping the continuous space to a discrete space, the dimension of the solution space of the GD association variable $\boldsymbol{X}$ may grow exponentially with the number of the GDs and the AAVs, making it difficult for DRL to train and converge. Moreover, some decision variables may encounter the problem of the varying action dimensions when converted into the elements of the action space. Specifically, when considering the bandwidth allocation and offloading decisions, the number of the GDs that need to be considered depends on the number of the GDs associated with the AAV in time slot $t$. Therefore, further consideration of the action space is required. These challenges motivate us to simplify and transform the action space, which is described as follows.

\subsubsection{Action Reduction and Transformation}

\par To ensure stable DRL training and convergence, an efficient low-complexity GD association strategy is first employed to reduce the action-space dimensionality. Specifically, GS-based methods, characterized by low complexity and stable matching, are widely used to obtain GD associations in each time slot. The GS-based GD association strategy is summarized as follows. Initially, GD $g$ sends a request to its nearest AAV $v$, and AAV $v$ accepts GD $g$ if it has not reached its service capacity. Otherwise, the AAV compares its currently served GDs and disconnects the farthest one to admit the new GD. The disconnected GD then requests the second-nearest AAV and repeats this process until all AAVs reach capacity or all GDs exhaust their available AAV choices. The detailed description is presented in Algorithm \ref{alg:association}.

\par Subsequently, the problems of the hybrid action space and varying action dimensions need to be addressed. Suppose that in time slot $t$, the number of GDs associated with AAV $v$ is $m$. For the discrete offloading decision variable $\boldsymbol{O}$, its relatively small dimension has minimal impact on the DRL algorithm training stability. As the offloading decisions for each GD are independent, we can define them as continuous actions and only consider the top $m$ actions based on the the numerical values for offloading actions of AAV $v$, which are then mapped to the discrete decisions. This approach eliminates the variable action dimension issue while maintaining the decision efficacy. For the variable dimensionality of bandwidth allocation, we implement a softmax-based mapping method since the total bandwidth for each AAV is fixed and the bandwidth of each associated GD needs to be determined. Specifically, we extract the $m$ highest values from the bandwidth allocation policy for AAV $v$, obtain proportions using the softmax function, and multiply these by the total bandwidth of each AAV to determine the allocated bandwidth for each associated GD, which can be given by

\begin{equation}
    B_{v,g}(t) =B_v \times \frac{\exp{(b_{v,g}(t))}}{\sum\nolimits_{g'\in\mathcal{G}_v(t)}\exp{(b_{v,g'}(t))}},
\end{equation}
where $\mathcal{G}_v(t)$ is the set of GDs associated with AAV $v$, $b_{v,g}(t)$ is the raw bandwidth allocation value for GD $g$ from the policy, and $B_v$ is the total bandwidth available for AAV $v$.

\begin{algorithm}[tbp]
    \small
    \caption{GS-based GD association strategy}
    \label{alg:association}
    \textbf{Initialize:} Create an available AAV list $\mathcal{V}_g$ and set variable $\mathrm{CONS}_g \leftarrow \mathrm{False}$ for $g \in \mathcal{G}$\;
    \While{$\exists g, \mathrm{CONS}_g = \mathrm{False}$ and $\mathcal{V}_g \ne \emptyset$}{
        Choose the nearest AAV $v \in \mathcal{V}_g$\;
        $X_{v,g}(t) \leftarrow 1$\;
        $\mathrm{CONS}_g \leftarrow \mathrm{True}$\;
        \If{$\sum_{g \in \mathcal{G}}X_{v,g}(t) > N_g^{max}$}{
        Find the farthest GD $g_f$\;
        $X_{v,g_f}(t) \leftarrow 0$\;
        $\mathcal{V}_{g_f} \leftarrow \mathcal{V}_{g_f} \backslash \{v\}$\;
        $\mathrm{CONS}_{g_f} \leftarrow \mathrm{False}$\;
        }
    }
    Return $X(t)$\;
\end{algorithm}

\par Hence, the high-dimensional hybrid variable-dimension action space derived from the decision variables can be reduced and transformed into a fixed-dimension continuous action space related to the AAV movement control, offloading decisions, and bandwidth allocation. Next, the specific MDP formulation is provided.

\subsubsection{MDP Formulation}

\par Benefiting from the aforementioned simplification and transformation of the action space, the considered optimization problem can be reformulated as an MDP with a continuous action space, and the key definitions are given as follows:

\par \begin{itemize} 
    \item \textit{State Space:} Since the task offloading and DC in the joint MEC-DC system depend on the relative distance of the AAVs and the GDs, the task status of GDs, the amount of stored data, and the remaining time, we define the states as follows:
\begin{equation}
    s_t=\{P_{\mathcal{V}}^t, P_{\mathcal{G}}^t,U_{\mathcal{G}}^t,D_{\mathcal{G}}^t,t_r\},
\end{equation}
where $P_{\mathcal{V}}^t$, $P_{\mathcal{G}}^t$, $U_{\mathcal{G}}^t$, and $D_{\mathcal{G}}^t$, $t_r$ represent the AAV positions, the GD positions, the urgency of the earliest tasks of GDs, the amount of stored data of GDs, and the number of remaining time slots in time slot $t$, respectively. Specifically, $U_{\mathcal{G}}^t$ equals the product of the deadline and the maximum tolerable delay of the earliest task of each GD in time slot $t$, and $t_r$ equals $T-t$.

    \item \textit{Action Space:} As aforementioned, in time slot $t$, the AAV movement, offloading decisions, and bandwidth allocation are optimized in the MDP to minimize the average E2E delay and AAV energy consumption while maximizing the amount of collected data. Consequently, the action set is defined as follows:
\begin{equation}
    a_t=\{m_{\mathcal{V}}^t,\alpha_{\mathcal{V}}^t,o_{\mathcal{V}}^t,b_{\mathcal{V}}^t\},
\end{equation}
where $m_{\mathcal{V}}^t$, $\alpha_{\mathcal{V}}^t$, $o_{\mathcal{V}}^t$, and $b_{\mathcal{V}}^t$ represent the distance, the direction of AAV movement, and the decisions related to offloading as well as the bandwidth allocation for each associated GDs of the AAVs in time slot $t$, respectively.

    \item \textit{Reward Function:} In DRL models, the immediate rewards obtained after the output action interacts with the environment provide a quantifiable measure of the policy in a given state, and subsequently, the agent continuously learns the optimal policy based on the returned rewards. Thus, the formulated joint optimization problem of minimizing the average E2E delay and AAV energy, as well as maximizing the volume of collected data, can be transformed into a problem of maximizing the cumulative rewards, and the reward function can be defined as follows:
\begin{equation}
    r_t = r_l^t+\rho_1r_d^t+\rho_2r_e^t+r_p^t,
\end{equation}
where $\rho_1$ and $\rho_2$ are two weighting factors used to unify the reward terms to the same order of magnitude. $r_l^t$ is equal to the difference between the maximum tolerable delay of the task and the actual E2E delay. Moreover, $r_d^t = D(t)$ and $r_e^t = -E(t)$ respectively quantify the total amount of collected data and the total energy consumption of all AAVs in time slot $t$. Additionally, $r_p^t$ is a penalty reward, applicable when an AAV violates the boundary constraints or collides with another AAV.
\end{itemize}

\begin{figure*}[tbp]
\centering
\includegraphics[width=0.99\textwidth]{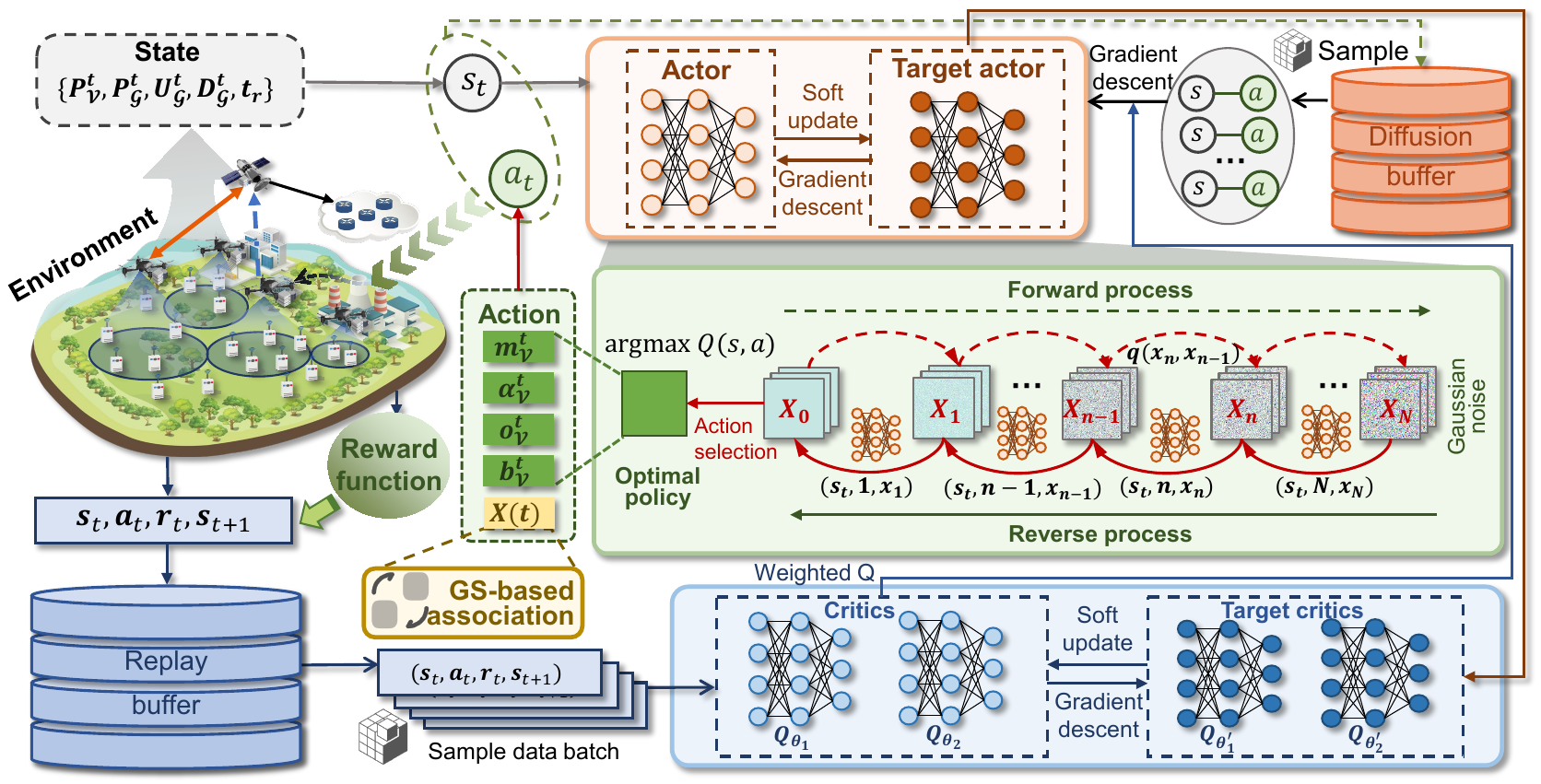}
\caption{The schematic of the proposed QAGOB approach. The architecture integrates a diffusion model into the DRL actor-critic structure to serve as the policy network, whereby the diffusion policy generates complex continuous actions via a reverse denoising process. Furthermore, the device association is decoupled from the neural network and efficiently solved by a GS-based approach.}
\label{fig:schematic}
\end{figure*}

\subsection{Diffusion Model-based Solution}

\par In this section, the motivation for using the diffusion model-based methods is first elaborated. Subsequently, the fundamental principles of the QVPO are introduced. Consequently, the QAGOB approach is proposed.

\subsubsection{Motivation for Using Diffusion Model-based Method}

\par DRL typically employs multilayer perceptrons (MLPs) as function approximators in the policy networks due to their simplicity and ease of implementation. However, when using MLPs to output the actions in the complex, high-dimensional continuous control scenarios, they suffer from the following limitations. First, MLP-based policies often produce the single-modal action distributions (\textit{e.g.}, Gaussian), which fail to capture the multiple viable strategies in the complex environments \cite{Du2024}. For instance, the agent may need to choose between mutually exclusive actions to satisfy MEC, DC, or energy-saving requirements. A single-modal output poorly represents such a scenario, as it cannot assign high probabilities to multiple separated action regions simultaneously. Second, MLPs inherently limit the exploration capabilities as their deterministic or the narrow stochastic sampling tends to converge prematurely to the suboptimal modes, especially in the sparse-reward or high-dimensional spaces \cite{du2024a}. Consequently, in the high-dimensional continuous control tasks, such as the considered satellite-AAV joint MEC-DC scenario, traditional MLP-based policy architectures may lead to inefficient learning and poor generalization, struggling to adapt to the high-dimensional continuous control scenarios.

\par In contrast, generative approaches offer a transformative approach to the policy representation in DRL, which can address the core limitations of the MLP-based methods. Among them, transformer-based and generative adversarial network (GAN)-based approaches can be integrated with DRL. However, they typically produce single-modal action distributions and suffer from unstable training processes \cite{du2024a,Paster2022}. Diffusion models, on the other hand, provide a promising alternative through their capability for multi-modal action generation, structured exploration, and progressively stable training characteristics \cite{Chi2025}. First, diffusion models inherently support multi-modal action distributions by modeling the policies as stochastic processes and enhance the representation of strategies. Second, the noise-adding and denoising mechanism of diffusion models enables the structured exploration. Specifically, by iteratively refining the actions through the reverse diffusion steps, they balance the exploration (stochastic sampling) and exploitation (gradient-guided optimization), thereby avoiding local optima, which is common in MLP-based policies. Furthermore, diffusion-based policies demonstrate superior training stability and robustness since their gradual denoising process enables smoothing policy optimization and mitigating gradient variance. By combining these properties, diffusion models achieve higher exploration performance, adaptability to the multi-modal tasks, and reliability in the complex, highly dynamic, and high-dimensional continuous control environments. Therefore, we employ a diffusion model-based enhanced DRL approach to solve the considered optimization problem.

\subsubsection{Preliminary of Denoising Diffusion Probabilistic Models (DDPM)}

\par Diffusion models such as DDPM \cite{ho2020}, are generative models that learn to gradually denoise data through a sequential process. They consist of two key phases, which are a forward process that incrementally corrupts data into noise, and a reverse process that learns to reverse this corruption to recover the original data distribution. In what follows, the detailed procedure of DDPM is provided.

\par \textit{Forward Process:} Given an original data sample $x_0$ , the forward process progressively adds the Gaussian noise over $N$ steps. At each step $n$, the noisy sample $x_n$ is generated from $x_{n-1}$ as follows:
\begin{equation}
    q(x_n|x_{n-1})=\mathcal{N}(x_n;\sqrt{1-\beta_n}x_{n-1},\beta_n\boldsymbol{I}),
\end{equation}
where $\boldsymbol{I}$ denotes the identity matrix, and $\beta_n$ is the variance schedule. By leveraging the Markov chain property, the posterior probability of $x_N$ conditioned on $x_0$ can be derived in a closed form as follows:
\begin{equation}
    q(x_N|x_0) =  \textstyle \prod_{n=1}^{n=N}q(x_n|x_{n-1}).
\end{equation}

\par Moreover, the relationship between $x_n$ and $x_0$ at any step $n$ in the forward process is described as follows:
\begin{equation}
    x_n=\sqrt{\bar{\alpha}}x_0 + \sqrt{1-\bar{\alpha}_n}\epsilon,
\end{equation}
where $\bar{\alpha}_n=\textstyle \prod_{n=1}^N\alpha_n$, wherein $\alpha_n=1-\beta_n$, and $\epsilon\sim\mathcal{N}(0,\boldsymbol{I})$ is a standard Gaussian noise. Eventually, $x_N$ converges to an isotropic Gaussian distribution $\mathcal{N}(0,\boldsymbol{I})$.

\par \textit{Reverse Process:} The reverse process starts from noise distribution $p(x_N)= \mathcal{N}(x_N;0,\boldsymbol{I})$ and iteratively denoises it to reconstruct $x_0$. At each step $n$, the model learns to approximate the true posterior $q(x_{n-1}|x_n)$ using a parameterized Gaussian transition:
\begin{equation}
    p(x_{n-1}|x_n)=\mathcal{N}(x_{n-1};\mu_{\theta}(x_n,n),\Sigma_{\theta}(x_n,n)),
\end{equation}
where $\mu_{\theta}$ and $\Sigma_{\theta}$ are the learnable mean function and variance function implemented as a neural network parameterized by $\theta$, respectively. As such, the reverse process is trained to reverse the forward diffusion steps, gradually refining $x_N$ into $x_0$.

\par \textit{Variational Lower Bound (VLB) Loss:} During the training process, the objective of DDPM is to maximize the VLB, which is given by $L_{VLB} = \mathbb{E}_{q(x_{0:N})}\left [ \log \frac{p_\theta(x_{0:N})}{q(x_{1:N}|x_0)} \right ]$ and can be simplified to the final loss, which is given by
\begin{equation}
    \mathbb{E}_{n\sim [1,N],x_0,\epsilon_n}\left [ \left \| \epsilon_n - \epsilon_\theta(\sqrt{\bar{\alpha_n}}x_0+\sqrt{1-\bar{\alpha}_n}\epsilon_n,n) \right \|^2 \right ].
\end{equation}

\subsubsection{Basic Principles of QVPO}

\par QVPO is a diffusion-based model-free online RL algorithm. Specifically, QVPO introduces several novel techniques to improve the sample efficiency and training stability of online RL, including the Q-weighted variational loss, entropy regularization, and efficient behavior policy design. These mechanisms address the key bottlenecks encountered when combining diffusion models with online RL, such as obtaining high-quality state-action sets as training samples, controlling training costs with effective exploration capabilities, and reducing the high variance of diffusion policies during the online interactions. By coordinating a diffusion-based policy network with value-guided optimization, QVPO is able to achieve robust learning in the complex high-dimensional continuous control environments.

\par \textit{Q-weighted Variational Objective for Diffusion Policy:} To train the diffusion policy using the variational bound objective, QVPO proposes a Q-weighted variational objective. Specifically, by adding the Q-values of the state-action pairs to the VLB objective, the diffusion policy can be effectively optimized. In this process, to address the issues of negative Q-values and the requirement for the high-quality training samples, QVPO further introduces a transformation function to convert Q-values into equivalent non-negative weights and obtain the high-quality samples, which is given by
\begin{align}
    &\mathcal{L}(\theta) \triangleq \mathbb{E}_{s, a \sim \pi_{k}(a \mid s), \epsilon, n}\Big[\omega_{e q}(s, a)\left\|\epsilon-\epsilon_{\theta}\left(\sqrt{\bar{\alpha}_{n}} a + \right.\right. \nonumber \\
    &\left.\left.\sqrt{1-\bar{\alpha}_{n}} \epsilon, s, n\right)\right\|^{2}\Big],\label{eq:lossa}\\
    &\omega_{e q}(s, a) \triangleq \omega_{qadv}(s, a) = \begin{cases} 
    A(s, a), & A(s, a) \geq 0, \\
    0, & A(s, a) < 0,
    \end{cases}\label{eq:weight}
\end{align}
where $A(s, a)=Q(s,a)-V(s,a)$ is the advantage function and $V(s,a)$ denotes the importance of the state $s$.

\par \textit{Diffusion Entropy Regularization:} To enhance the exploration capability of the diffusion model within limited steps to avoid the excessive training and evaluation costs, QVPO introduces an entropy regularization term for the diffusion policy as follows:
\begin{equation}\label{eq:losse}
\begin{aligned}[t]
    \mathcal{L}_{e n t}(\theta) \triangleq \mathbb{E}_{s, a \sim \mathcal{U}, \epsilon, n}\left[\omega_{e n t}(s)\left\|\epsilon-\boldsymbol{\epsilon}_{\theta}\left(\sqrt{\bar{\alpha}_{n}} a+\right.\right.\right.\\
    \left.\left.\left.\sqrt{1-\bar{\alpha}_{n}} \boldsymbol{\epsilon},s, n\right)\right\|^{2}\right],
\end{aligned}
\end{equation}
where $\omega_{ent}(s)=\omega_{ent}\sum\nolimits_{i=1}^{N_s}\frac{\omega_{eq}(s,a_i)}{N_s}$ is a coefficient related to state $s$ for balancing the exploration and exploitation and $N_s$ is the number of training samples selected from the diffusion strategy for state $s$.

\par \textit{Reducing Diffusion Policy Variance via Action Selection:} Despite the diffusion model enabling online RL agents to explore diverse policies, its inherent stochasticity introduces significant policy variance, leading to inefficiency in the environment interactions via the behavior policy. To mitigate this, QVPO proposes an efficient behavior policy $\pi_{\theta}^{\varsigma}$, which enhances the sample efficiency by prioritizing high-Q actions during the exploration. Specifically, the core idea is that the action samples with higher Q-values are more likely to produce trajectories with substantial cumulative rewards, thus being more valuable for training. Formally, the efficient behavior policy is defined as follows:
\begin{equation}\label{eq:action selection}
    \pi_{\theta}^{\varsigma}(a \mid s) \triangleq \underset{a \in\left\{a_{1}, \ldots, a_{\varsigma} \sim \pi_{\theta}(a \mid s)\right\}}{\operatorname{argmax}} Q(s, a),
\end{equation}
where $\varsigma$ represents the action selection number. Note that due to the significant difference in the diffusion policy, a value smaller than $\varsigma$ can be used when calculating the target policy for the TD target to avoid the severe overestimation of the target Q-values.

\begin{algorithm}[tb]
    \small
    \caption{QAGOB}
    \label{alg:QAGOB}
    \textbf{Initialize:} Replay buffer $\boldsymbol{R_b}$, diffusion buffer $\boldsymbol{R_d}$, diffusion-based policy network $\pi_{\phi}$ with parameter $\phi$, critic networks $Q_{\theta_1}$, and $Q_{\theta_2}$ with parameter $\theta_1$, $\theta_2$\;
    \For{episode$=1,2,\cdots, E$}{
    Reset the environment and observe the initial state $s_0$\;
    \For{$t=1,2,\cdots, T$}{
        Sample $\varsigma$ candidate actions $\{a_i\}_{i=1}^{\varsigma}\sim \pi_{\phi}$ through the denoising process\;
        Select the action $a_t$ with the highest Q value according to Eq. \eqref{eq:action selection}\;
        Store state-action pair $\{s_t,a_t\}\to \boldsymbol{R_d}$\;
        Calculate GD association matrix $\boldsymbol{X(t)}$ by using \textbf{Algorithm \ref{alg:association}}\;
        Apply action $a_t$ and association matrix $\boldsymbol{X(t)}$ to obtain reward $r_t$, and observe next state $s_{t+1}$\;
        Store transition $(s_t, a_t, r_t, s_{t+1}) \to \boldsymbol{R_b}$\;
        Sample $N_{\pi}$ state-action pairs from $\boldsymbol{R_d}$, and generate $N_u$ samples from uniform distribution $\mathcal{U}$\;
        Endow the $N_{\pi}$ samples with the weights according to Eq. \eqref{eq:weight}\;
        Select an action $a_{max}$ with the maximum weight among $N_{\pi}$ samples\;
        Endow the $N_u$ samples with weight $\omega_{ent}(s)=\omega_{ent} \cdot \omega_{eq}(s,a_{max})$\;
        Update the policy network parameters by minimizing the summation of Eqs. \eqref{eq:lossa} and \eqref{eq:losse}\;
        Sample a batch of transitions from $\boldsymbol{R_b}$\;
        Construct the TD target as $y_t=r_t+\gamma Q_{\theta}(s_{t+1},$\
        $\pi_{\phi}^{\varsigma}(a|s_{t+1}))$ for each transition\;
        Update the critic network parameters via MSE loss \cite{Ding2024}\;
        Update the target network parameters:\
        $\overline{\theta}_i \leftarrow \varrho \theta_i + (1-\varrho)\overline{\theta}_i$, $i \in \{1,2\}$,\
        $\overline{\phi} \leftarrow \varrho \phi + (1-\varrho)\overline{\phi}$,\
        where $\varrho$ is the soft update parameter.
    }
    }
\end{algorithm}

\subsection{The Proposed QAGOB Methodology}

\par In this section, the workflow of the proposed QAGOB solution is presented, followed by the computational complexity analysis of the proposed QAGOB approach.

\subsubsection{Main Flow of QAGOB}

\par Building on the formulated MDP framework, the GS-based GD association strategy, and the QVPO algorithm, we propose the QAGOB method to tackle the joint optimization problem in the considered satellite-AAV-enabled joint MEC-DC system. Fig. \ref{fig:schematic} illustrates the schematic of the proposed QAGOB method. the GS-based GD association strategy is embedded within the RL framework, serving as a dedicated module for handling discrete device association. This module cooperates with the diffusion model to interact with the environment and compute a unified reward. Ultimately, this design overcomes critical challenges in the MDP formulation, such as the design of the hybrid action space. Moreover, by leveraging the intrinsic multi-modal exploration capabilities of diffusion models \cite{du2024a}, the capability of online RL algorithms to navigate the optimal decisions in high-dimensional continuous control environments can be enhanced.

\par The detailed steps of the proposed QAGOB approach are elaborated in Algorithm \ref{alg:QAGOB}. Specifically, during each time slot, the algorithm first observes the initial state and samples $\varsigma$ candidate the actions from the policy network via its denoising process, selects action $a_t$ with the highest Q-value using Eq. \eqref{eq:action selection}, and stores the state-action pair in $\boldsymbol{R_d}$. Concurrently, GD association matrix $X(t)$ is calculated via Algorithm \ref{alg:association} and is applied with $a_t$ to interact with the environment, observes the reward $r_t$ and next state $s_{t+1}$, and stores the transition $(s_t, a_t, r_t, s_{t+1})$. For policy updates, the QAGOB samples $N_{\pi}$ weighted state-action pairs from $\boldsymbol{R_d}$ and $N_u$ uniform exploration samples, selects the highest-weight action $a_{max}$, and optimizes the policy network by minimizing the combined loss from Eqs. \eqref{eq:lossa} and \eqref{eq:losse}. The critic networks are updated using the TD targets derived from the diffusion policy, while the target networks are softly updated to stabilize training.

\subsubsection{Computational Complexity Analysis}

\par In this part, we analyze the computational complexity of the QAGOB during the training and execution phases.

\par \textit{Training Phase:} The computational complexity of the training process mainly consists of two parts, which are the action sampling and network update, respectively. 
\begin{itemize}
    \item Action Sampling: During the action sampling phase, the actor network generates actions via the reverse diffusion in each time slot, and the complexity of interacting with the environment during training phase is $\mathcal{O}(ETV)$, where $E$ denotes the number of episodes, $T$ represents the number of time slots per episode, and $V$ is the complexity of interacting with environment. Furthermore, the complexity of generating actions using the diffusion-based actor network is $\mathcal{O}(ETN|\phi|)$, where $N$ is the number of denoising steps, and $|\phi|$ is the number of parameters in the actor network. Consequently, the complexity in the action sampling part is $\mathcal{O}(ET(V+N|\phi|))$~\cite{Liang2025, Du2024}.
    \item Network Parameter Update: For the network update phase, the computational complexity mainly comes from the parameter updates of the actor network, the two critic networks, and their respective target networks. Since the parameters of the target network are of the same as the original network, the computational complexity of this phase is $\mathcal{O}(2ET/d|\phi|+2ET(|\theta_1|+|\theta_2|))$~\cite{Zhang2025b}, where $d$ represents the update frequency of the actor network, $|\theta_1|$ and $|\theta_2|$ denotes the number of parameters of the two critic networks, respectively.
\end{itemize}
\par Therefore, the total computational complexity of the training phase is $\mathcal{O}(ET(V+(N+2/d)|\phi|+2(|\theta_1|+|\theta_2|))$.

\par \textit{Execution Phase:} In the execution phase, the computational complexity mainly stems from the back propagation of the diffusion-based actor network generating the actions according to the current state. Thus, the computational complexity of the execution phase is $\mathcal{O}(ETN|\phi|)$ \cite{zhang2025}. This complexity is significantly lower than that during the online training phase, making it feasible to deploy the algorithm on AAVs or other mobile computing units.

%
% Simulation And Analyses
%

\section{Simulation and Analysis}
\label{sec:experiments_and_analysis}

\par In this section, we evaluate the performance of the proposed QAGOB algorithm in addressing the considered optimization problem and verify its effectiveness under different settings.

\subsection{Simulation Configuration}

\par In this section, the detailed descriptions of the simulation platform, system parameter settings, and benchmarks are provided. In addition, the details about the components of the algorithm can be found in Appendix B.

\subsubsection{Simulation Platform}

\par Our simulations are conducted on a workstation equipped with an NVIDIA GeForce RTX 3090 GPU with 24 GB of memory and a 32-core 13th Gen Intel(R) Core(TM) i9-13900K processor with 128 GB of RAM. The operating system on the workstation is Ubuntu 22.04.3 LTS, and we utilize PyTorch 2.2.1 with CUDA 11.8 for all the DRL-based algorithms.

\subsubsection{System Parameter Settings}

\par In our simulation, a square area of size $3000 \times 3000$ m\textsuperscript{2} is considered, where $(x_{min}, y_{min})=(-1500,-1500)$ and $(x_{max},y_{max})=(1500,1500)$. The numbers of the AAVs and GDs are 4 and 30, respectively, and the maximum number of the served GDs of each AAV is $N_g^{max}=4$. Furthermore, the locations of the GDs are randomly distributed within the area, while the initial positions of the AAVs are set to $(-750, -750)$, $(-750, 750)$, $(750, -750)$, and $(750, 750)$, respectively, in order to prevent the severe interference and collision tendencies at the beginning. Moreover, the altitudes of the AAVs and the satellite are 100 m and 800 km \cite{huang2024}, respectively. In addition, the safe distance between the AAVs is 50 m, and the maximum moving speed of an AAV is 50 m/s \cite{zhan2021}. For MEC, the task size follows a Poisson distribution with the parameter $\lambda_{MEC}$, namely $L_{g,k} \sim P(\lambda_{MEC}) \times 10^5$ bits. Similarly, for the DC, the generated data size in time slot $t$ is $D_g(t) \sim P(\lambda_{DC})\times10^4$ bits. Besides, the remaining parameter settings and training configurations are presented in Tables~\ref{tab:parameter} and \ref{tab:network}, respectively.

\begin{table}[tbp]
\caption{Simulation parameters}
\label{tab:parameter}
\setlength{\tabcolsep}{8pt}
\renewcommand{\arraystretch}{0.8}
\begin{tabular}{@{}ll@{}}
\toprule
Parameters                  & Value                                            \\ \midrule
Task density coefficient $\delta_f$         & 0.1      \\
Poisson distribution rate $\lambda_{MEC}$, $\lambda_{DC}$        & 6, 10 \\
Length of deadline for each task $D_{m,f}$         & [10, 30] s \\
Max tolerance time limit $T_f^{max}$              & [0.75, 1.75] s \\
Noise power spectrum density $n_0$          & $-174$ dBm/Hz \cite{gao2024}               \\
Excessive propagation losses $\eta_{LoS}$, $\eta_{NloS}$ & 0.1, 21 \cite{chen2023}           \\
Transmit power $p_g^{max}$, $p_v^{max}$, $p_s^{max}$           & 0.3 \cite{zhang2024}, 0.5 \cite{zhao2025}, 20 \cite{xu2023}  W \\
Antenna gains $G_v$, $G_s$         & $10^5$, $10^5$ \cite{xu2023} \\
Transmission rate threshold $R_{th}$   & 1 Mbps\\
Bandwidth $B_v$, $B_s$           & 5 MHz \cite{chen2025}, 1 MHz \cite{zhang2023}                                            \\
Required CPU cycles per bit data $C$                & 1000 cycles/bit   \\
Effective switching capacitance $\kappa$   & $8.2 \times10^{-9}$ J/cycle \cite{jiang2023}    \\

CPU computing capability $\omega_v$, $\omega_s$  & 8, 20 GHz \cite{zhu2025}        \\
Rain attenuation $F_{rain}$    & 6 dB \cite{xu2023, qin2024}\\ \bottomrule
\end{tabular}%
\end{table}

\begin{table}[tbp]
\centering
\caption{Training configurations}
\label{tab:network}
\setlength{\tabcolsep}{28pt}
\renewcommand{\arraystretch}{0.8}
\begin{tabular}{@{}lll@{}}
\toprule
Parameters                  & Value                                          &  \\ \midrule
Denoising steps for actor $N_g^{max}$    & 10                                &  \\
Network structure for critic   & [256, 128]                                        &  \\
Total episodes $E$                 & 3000                                          &  \\
Time step in each episode $T$     & 300                               &  \\
Discount factor $\gamma$       & 0.9                                          &  \\
Target soft update coefficient $\tau$     & 0.005       &  \\
Learning rate for actor        & $3\times 10^{-4}$                                         &  \\
Learning rate for critic       & $3 \times 10^{-2}$                                   &  \\
Replay buffer size             & $10^6$                                        &  \\
Entropy regularization coefficient     & 0.02                                           &  \\
Batch size                      & 256                                            &  \\
\bottomrule
\end{tabular}%
\end{table}

\subsubsection{Benchmarks}

\par To evaluate the performance of the QAGOB in addressing the considered optimization problem, we compare it with the following approaches:
\begin{itemize}
    \item \textit{Random Strategy:}~The random strategy determines the random AAV movement distances and directions, offloading decisions, and bandwidth allocation actions in each time slot. Additionally, the AAV-GD association matrix in each time slot is assigned random values within a reasonable range. This strategy is employed as a benchmark to measure the performance of other methods.
    \item \textit{Greedy Strategy:}~The greedy strategy selects the nearest GD for the AAV to approach during its movement and prioritizes associating with the nearest GD within the allowable service capacity range. Moreover, the offloading decisions and bandwidth allocation actions are randomly sampled through a uniform distribution.
    \item \textit{Proximal Policy Optimization (PPO)} \cite{ppo2017}: PPO is an efficient DRL algorithm, which constrains the magnitude of policy updates to enhance the stability. Moreover, PPO employs online learning, generating data directly from the current policy without the need for experience replay, thereby reducing computational complexity.
    \item \textit{Soft Actor-Critic (SAC)} \cite{sac2018}: SAC is a DRL algorithm based on the maximum entropy framework, which introduces an entropy regularization term to encourage the policy diversity, thereby promoting the exploration. Furthermore, SAC uses a twin Q-network to reduce the overestimation of Q-values and automatically adjusts the target entropy to avoid the excessive or insufficient exploration.
    \item \textit{Diffusion Model-Enhanced Twin-Delayed Deep Deterministic (DM-TD3)} \cite{Liang2025}: This paper proposes a DM-TD3 algorithm, which replaces the traditional MLP structure in the actor network of the original TD3 algorithm with a diffusion model to enhance the action exploration capability of the original TD3 algorithm, thereby optimizing the AAV trajectory, GD scheduling, and proportion of data transmission duration in order to minimize the age of information (AoI) of the GDs and the energy consumption of the AAVs.
\end{itemize}

\begin{figure*}[tbp]
\centering
\includegraphics[width=0.99\textwidth]{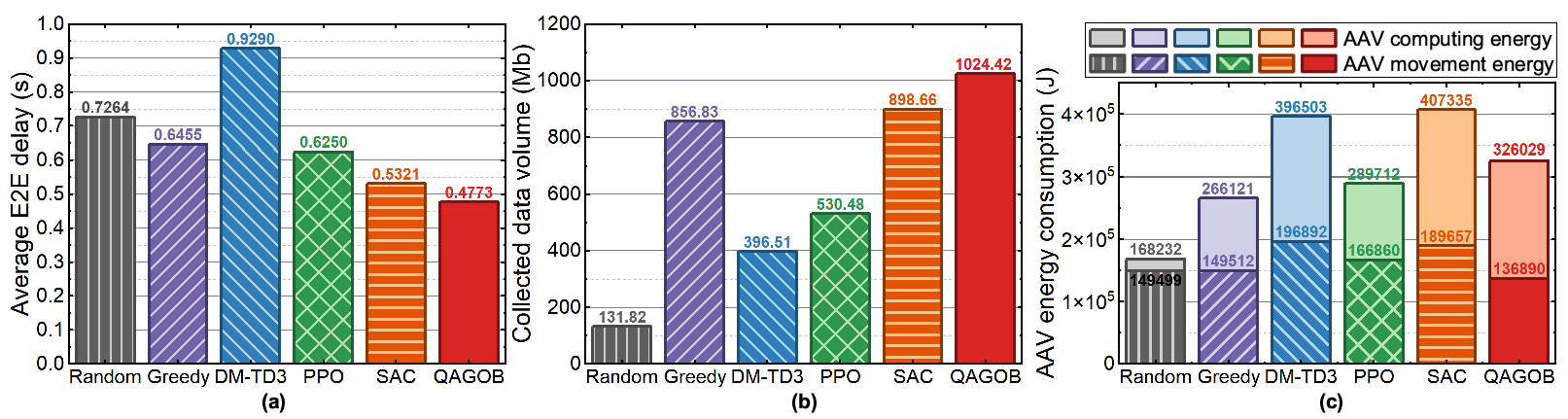}
\caption{Comparison results of QAGOB approach and benchmarks. (a) Average E2E delay. (b) Collected data volume. (c) AAV energy consumption.}
\label{fig:comparison}
\end{figure*}

\begin{figure}[tbp]
\centering
\includegraphics[width=0.49\textwidth]{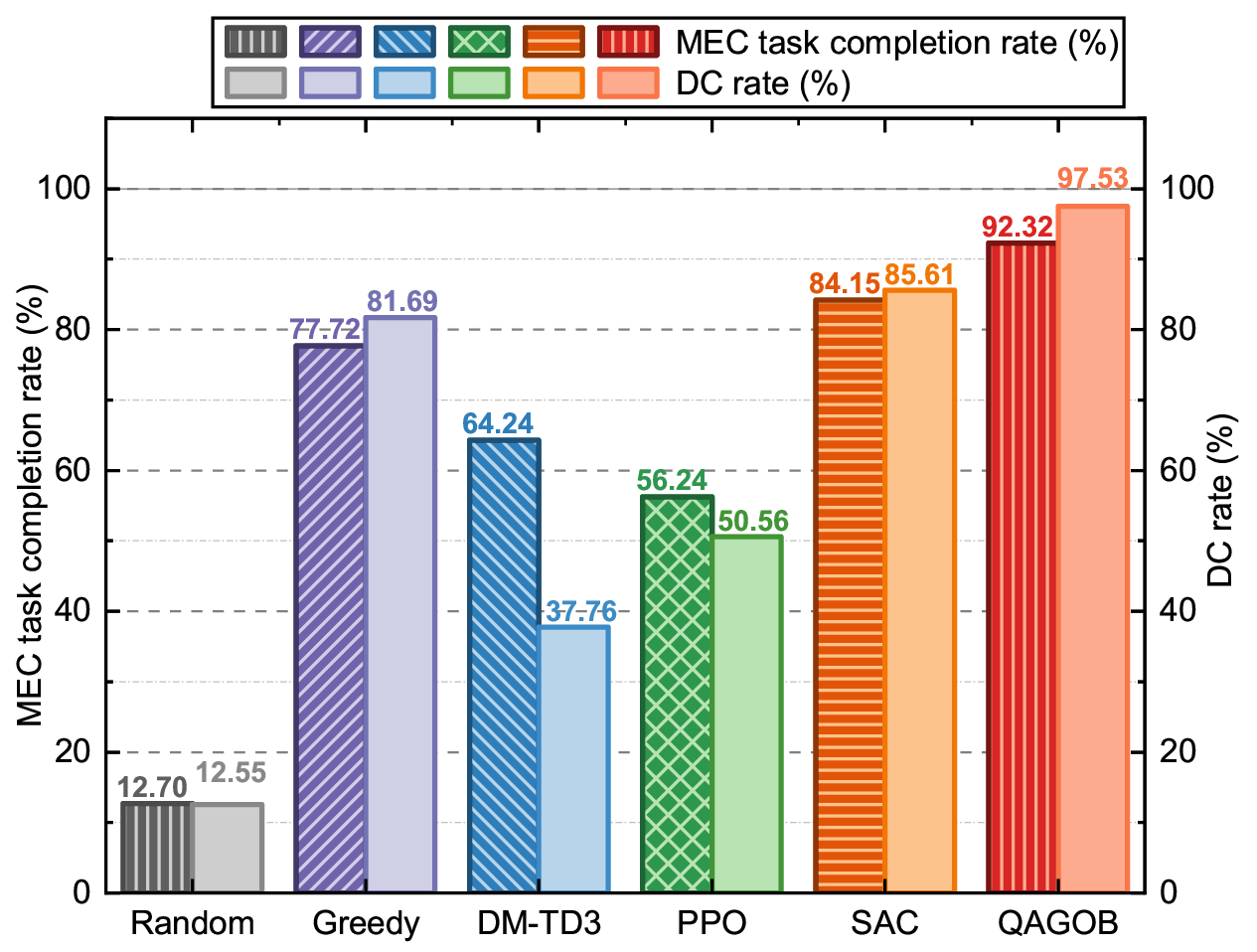}
\caption{Comparison results of MEC task completion rate and DC rate of QAGOB approach and benchmarks.}
\label{fig:completion rate}
\end{figure}

\subsection{Simulation Results}

\par This section provides the simulation results of the proposed QAGOB approach as well as the above-mentioned benchmark methods, including the comparison results, convergence results, AAV trajectory results, and performance analysis of the QAGOB under the different environment configurations and algorithm parameters. In addition, we provide comparative results for satellite and GD energy consumption, as well as comparative results with separately optimized MEC and DC, which can be found in Appendix C.1 and Appendix C.2, respectively.

\subsubsection{Comparison Results}

\par This part compares the performance of the proposed QAGOB approach with other methods in addressing the considered optimization problem. Specifically, Figs. \hyperref[fig:comparison]{3(a)}, \hyperref[fig:comparison]{(b)}, and \hyperref[fig:comparison]{(c)} respectively illustrate the average E2E delay of the system, the amount of collected data, and the energy consumption of AAVs. In addition, Fig. \ref{fig:completion rate} shows the performance of the system in terms of the MEC task completion rate and DC rate.

\par As shown in Fig. \hyperref[fig:comparison]{3(a)}, the proposed QAGOB method achieves the lowest average E2E delay. This result demonstrates the effectiveness of our proposed method in ensuring MEC performance in terms of delay. Additionally, we observe that the DM-TD3 method performs worse than the random strategy in addressing the considered optimization problem. This may be because directly replacing the actor network with a diffusion model may not be stable enough when dealing with different problems, which may face the challenges in learning the high-quality actions and exploration performance, thus being prone to falling into the local optimum.

\par As can be observed from Fig. \hyperref[fig:comparison]{3(b)}, the proposed QAGOB achieves superior data collection performance, accumulating 13.99\% more stored data than the second-best method within the specified time. While the greedy baseline demonstrates competent performance through the nearest-GD selection strategy, both SAC and QAGOB show better performance. This may be because SAC employs a maximum entropy strategy, which can explore better policies. Meanwhile, the QAGOB can not only facilitate exploration based on the entropy-regularized strategy but also learns better policies from the high-quality actions from the Q-weighted policies. As a result, the QAGOB can collect more stored data while ensuring the MEC performance.

\par From Fig. \hyperref[fig:comparison]{3(c)}, it can be seen that the proposed QAGOB approach achieves lower AAV energy consumption than the SAC and DM-TD3 methods after optimization (80.04\% and 82.23\% of their values, respectively), and the QAGOB obtains the lowest AAV movement energy consumption. Although the PPO, greedy, and random methods result in lower AAV energy consumption than that achieved by the QAGOB approach, these methods are significantly worse than the QAGOB approach in terms of the average E2E delay and the amount of collected data. Moreover, the second-lowest AAV movement energy consumption is 9.2\% higher than that of the QAGOB approach. This may be because the QAGOB approach, when considering the trade-off between the offloading computing tasks to the satellite to save the AAV computing energy and performing computations on the AAV to reduce delay, chooses to perform most of the MEC tasks on the AAV and minimizes the AAV movement energy consumption to ensure the joint optimization of the MEC performance of the system and the AAV energy consumption.

\par Additionally, the results in Fig. \ref{fig:completion rate} demonstrate the effectiveness of the proposed QAGOB and benchmark algorithms in terms of the MEC task completion rate and DC rate. It is evident that the QAGOB method achieves the highest MEC task completion rate and DC rate, showcasing its superior performance, which optimizes the average E2E delay and the amount of data collected while covering the vast majority of MEC tasks and DC data. In conclusion, the proposed QAGOB approach demonstrates superior performance in jointly optimizing the average E2E delay, the amount of collected data, and the AAV energy consumption, outperforming the benchmark algorithms.

\begin{figure}[t]
\centering
\includegraphics[width=0.49\textwidth]{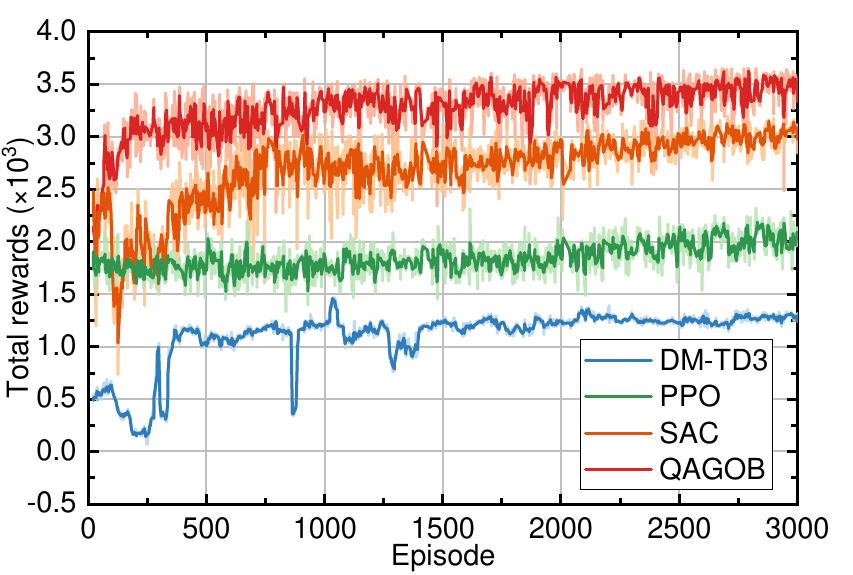}
\caption{Reward curves of QAGOB approach and benchmarks.}
\label{fig:converge}
\end{figure}

\subsubsection{Converge Results}

\par In this part, the convergence performance of the QAGOB approach is investigated, along with the analysis of the convergence results of the QAGOB method and other benchmarks.

\par Fig. \ref{fig:converge} illustrates the convergence performance of the QAGOB method compared with other DRL algorithms and diffusion model-based algorithms. It can be observed that the reward values obtained by the QAGOB method converge more rapidly and exhibit a consistently upward trend, with higher total rewards than that of the other benchmarks, demonstrating superior convergence performance. This may be attributed to the fact that the proposed algorithm can quickly learn the effective strategies through the Q-weighted high-quality actions and the multi-modal policy investigation in the early stage, and continuously explore better actions based on the inherent exploration capability of diffusion models and the entropy regularization term in the proposed approach.

\subsubsection{AAV Trajectory Results}

\par
\begin{figure}[t]
\centering
\includegraphics[width=0.70\linewidth]{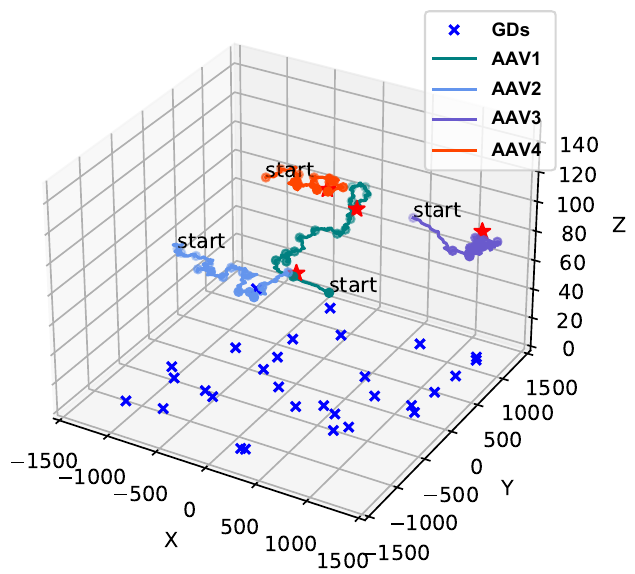}
\caption{AAV movement trajectory (The start and end points are marked with text and red stars, respectively).}
\label{fig:trajectory}
\end{figure}

 This part represents the AAV trajectory optimization result during the training of the QAGOB approach. It can be seen from the Fig. \ref{fig:trajectory} that all AAVs move within the specified area and near the locations where the GDs are clustered. Moreover, the four AAVs adopt a strategy of keeping a distance from other AAVs to avoid collisions as well as severe interference. This may be due to the algorithm has learned effective strategies from the environment and the proposed reward functions to find a balance among the three optimization objectives, which can not only ensure a high MEC and DC service coverage rate and the service performance for MEC and DC services but also saves a significant amount of the propulsion energy since long-distance movements are not required.

\subsubsection{Impact of Parameter and Environmental Settings}

\par In this part, we evaluate the performance of the QAGOB under the different algorithm parameters and environmental settings, including the effects of the number of the denoising steps and the maximum service capacity of the AAVs.

\par
\begin{figure}[tbp]
\centering
\includegraphics[width=0.49\textwidth]{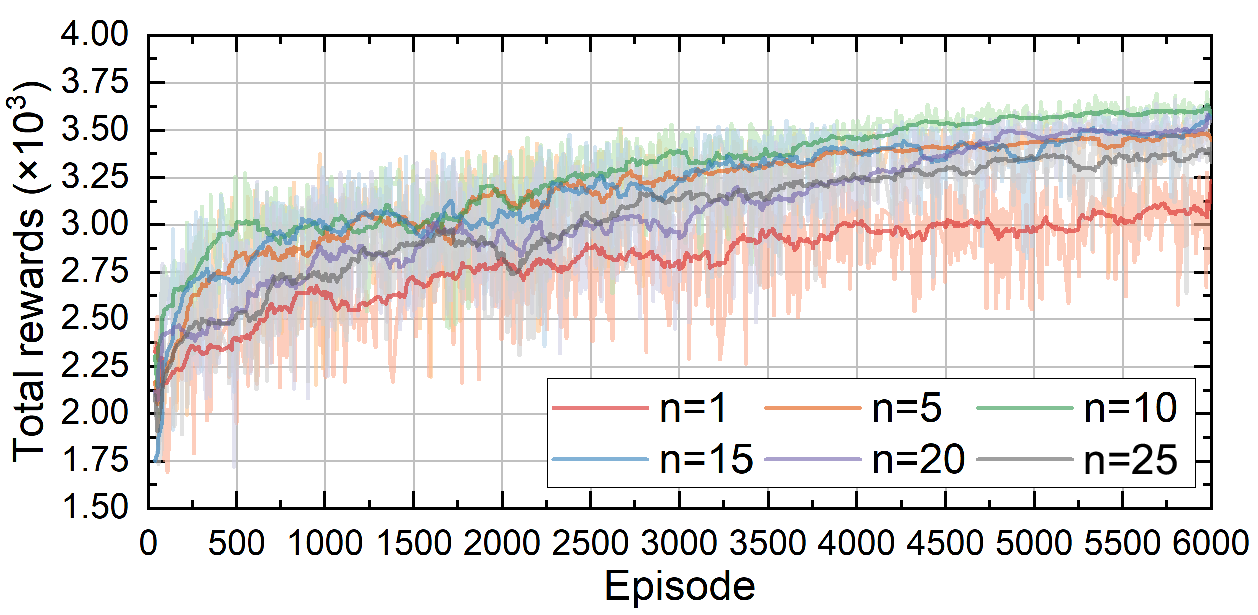}
\caption{Comparison results of reward curves of QAGOB with different denoising steps.}
\label{fig:n_timesteps}
\end{figure}

 \textit{Effect of Denoising Steps:} The number of denoising steps is a crucial parameter in diffusion models and can significantly affect the performance of the results. Specifically, a sufficient number of denoising steps can effectively process noise and generate high-quality samples. However, excessive denoising steps will increase the training time \cite{zhang2025}. Therefore, it is necessary to compare the performance of the QAGOB method under the different numbers of the denoising steps and select the appropriate parameters. As can be seen from Fig. \ref{fig:n_timesteps}, increasing the number of denoising steps initially improves algorithm performance, but subsequently leads to degradation, with optimal performance achieved when $n=10$. Notably, the worst performance occurs when $n=1$ and $n=25$. This may be because too few denoising steps do not provide sufficient capability to generate effective solutions, whereas excessive steps lead to overfitting of the noise patterns, thereby degrading the quality of the generated actions.

\par
\begin{figure}[tbp]
\centering
\includegraphics[width=0.49\textwidth]{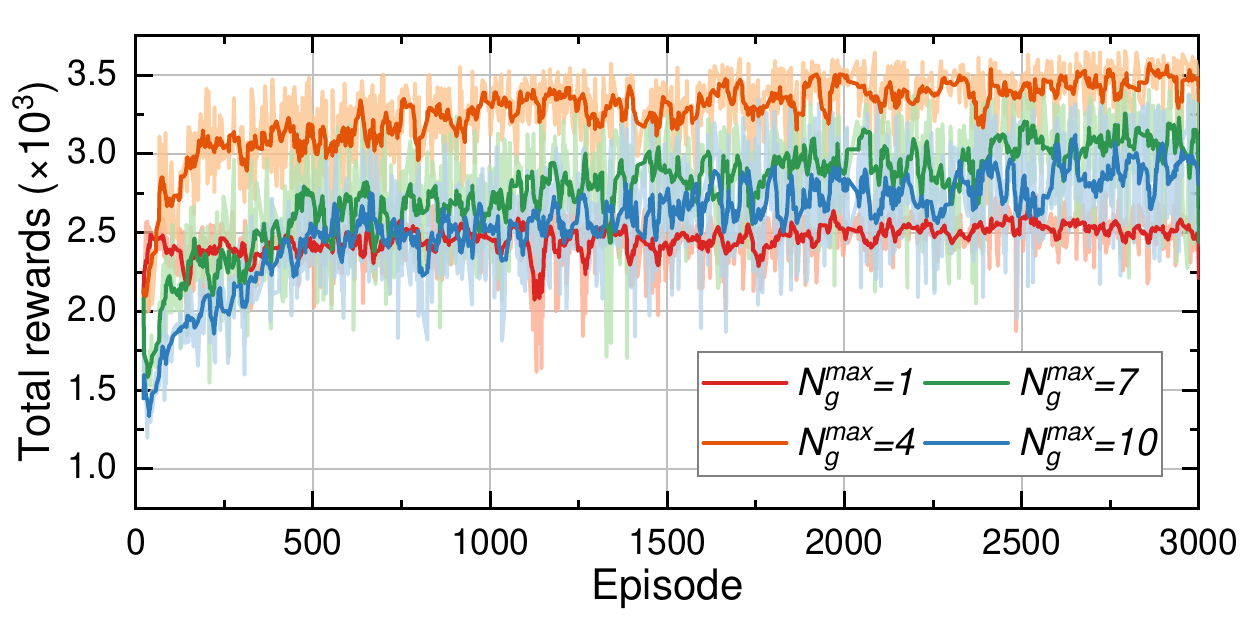}
\caption{Comparison results of reward curves of QAGOB with different maximum service capacity of AAVs.}
\label{fig:max_users}
\end{figure}

 \textit{Effect of Maximum Service Capacity of AAVs:} We evaluate the impact of the maximum AAV service capacity $N_g^{\max}$ on the QAGOB approach, as $N_g^{\max}$ inherently involves a trade-off between inter-cell interference and resource utilization. As shown in Fig. \ref{fig:max_users}, the reward first increases and then decreases with $N_g^{\max}$, peaking significantly at $N_g^{\max}=4$. This trend is attributed to several factors. Specifically, when $N_g^{max}$ is small, the AAV–GD channels suffer almost no co-channel interference, and the data rate is high. Thus, both task-completion and data-collection ratios are relatively large. However, most GDs remain unserved, and thus the total reward are very low. Conversely, as $N_g^{max}$ grows, each AAV can serve more GDs, and resource utilization improves. However, when $N_g^{max}$ exceeds a certain threshold, the excessive number of GDs within the same AAV coverage area causes severe co-channel interference between GDs, resulting in a significant drop in the effective communication rate of each GD. This directly degrades overall system performance. Consequently, $N_g^{max}=4$ offers the best trade-off between system throughput and service quality.
% As can be seen from Fig. \ref{fig:max_users}, the reward obtained by the QAGOB method is the highest when $N_g^{max}=4$ and the worst when $N_g^{max}=1$. Moreover, the reward declines as $N_g^{max}$ increases when $N_g^{max}>4$. This may be because around $N_g^{max}=4$, the AAVs can achieve the optimal balance between reducing interference and improving the task rewards, thereby obtaining higher rewards. Therefore, we set $N_g^{max}=4$ in the simulation parameter settings.

%
% Conclution
%
\section{Conclusion}
\label{sec:conclusion}

\par This paper has investigated a satellite-AAV-assisted joint MEC-DC system and has formulated a joint optimization problem to simultaneously minimize MEC delay and AAV energy consumption while maximizing collected data. We have reformulated the problem as an MDP with a transformed action space and have proposed the QAGOB method. This approach has been designed to leverage diffusion models to achieve real-time, feasible policies with multi-modal exploration and enhanced training stability. Simulation results have demonstrated that QAGOB has significantly outperformed DRL and diffusion-based benchmarks, achieving near-optimal results with high convergence robustness and efficiency. Additionally, while improving the energy consumption of the AAVs, QAGOB can achieve an 11.48\% reduction in MEC delay and a 13.99\% increase in the amount of collected data. Compared to optimizing the MEC and DC separately, the joint optimization reduces MEC latency by 10.13\%, increases the amount of the collected data by 28.93\%, and reduces the AAV movement energy consumption by 4.65\%.

% \par This paper has investigated a satellite-AAV-assisted joint MEC-DC system. Specifically, we have considered a joint MEC-DC scenario where a satellite and a set of AAVs act as MEC servers and DC nodes to collaboratively serve the GDs and have formulated a joint optimization problem to simultaneously minimize the MEC E2E delay and AAV energy consumption while maximizing the amount of collected data. The optimization problem has been reformulated as an MDP with a transformed action space to facilitate the use of DRL methods. Subsequently, we have proposed the QAGOB method to leverage the multi-modal policy generation, efficient exploration, and stable training capabilities of diffusion models to obtain the real-time feasible policies. Simulation results have demonstrated that the proposed QAGOB method outperforms other DRL benchmark algorithms and diffusion-based DRL benchmark algorithms. Furthermore, the training stability and convergence robustness of the proposed QAGOB method have been verified under various stochastic environments and parameter settings and we have found the proposed QAGOB method can achieve near-optimal results without requiring a high number of denoising steps. Additionally, while improving the energy consumption of the AAVs, QAGOB can achieve an 11.48\% reduction in MEC delay and a 13.99\% increase in the amount of collected data. Compared to optimizing the MEC and DC separately, the joint optimization reduces MEC latency by 10.13\%, increases the amount of the collected data by 28.93\%, and reduces the AAV movement energy consumption by 4.65\%.}

\bibliographystyle{IEEEtran}
\bibliography{bib}

%\newpage
\vspace{-10 mm}
\begin{IEEEbiography}[{\includegraphics[width=1in,height=1.25in,clip,keepaspectratio]{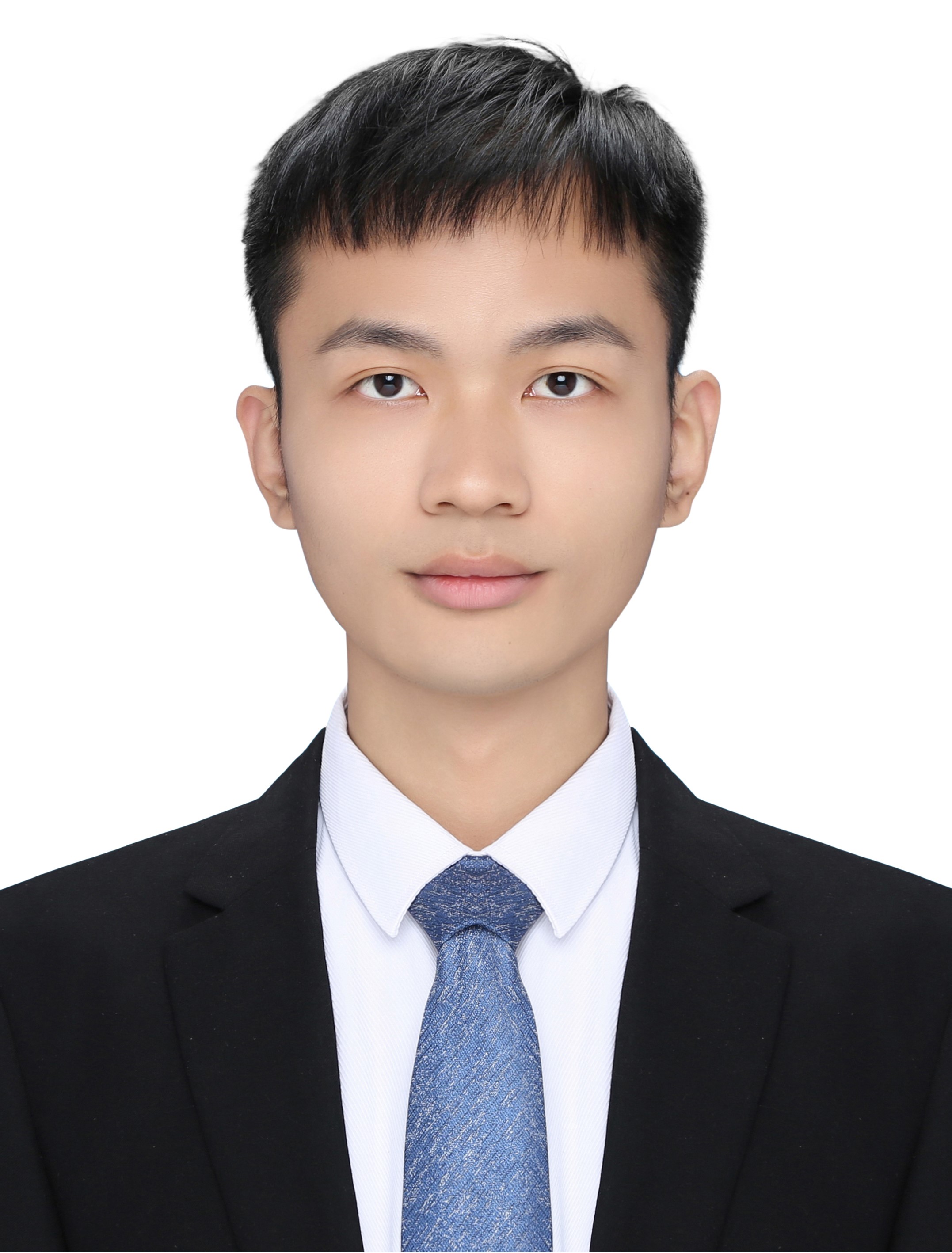}}]{Boxiong Wang}
received the B.S. and M.S. degrees in Software Engineering from Jilin University, in 2021 and 2024, respectively. He is currently pursuing the Ph.D. degree in the College of Computer Science and Technology at Jilin University, and is also a joint Ph.D. student at the Beijing Zhongguancun Academy. His current research focuses on AAV networks, mobile edge computing, and optimization.
\end{IEEEbiography}
\vspace{-13 mm}

\begin{IEEEbiography}[{\includegraphics[width=1in,height=1.25in,clip,keepaspectratio]{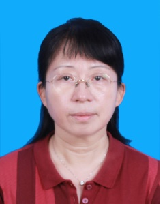}}]{Hui Kang}
received the M.E. and Ph.D. degrees from Jilin University in 1996 and 2007, respectively. She is currently a Professor with the College of Computer Science and Technology, Jilin University. Her research interests include grid computing, information integration, and distributed computing.
\end{IEEEbiography}
\vspace{-15 mm}

\begin{IEEEbiography}[{\includegraphics[width=1in,height=1.25in,clip,keepaspectratio]{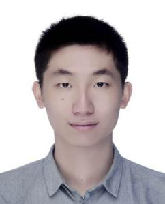}}]{Jiahui Li}
received his B.S. in Software Engineering, and M.S. and Ph.D. in Computer Science and Technology from Jilin University, Changchun, China, in 2018, 2021, and 2024, respectively. He was a visiting Ph.D. student at the Singapore University of Technology and Design (SUTD). He currently serves as an assistant researcher in the College of Computer Science and Technology at Jilin University. His current research focuses on integrated air-ground networks, UAV networks, wireless energy transfer, and optimization.
\end{IEEEbiography}
\vspace{-13 mm}

\begin{IEEEbiography}[{\includegraphics[width=1in,height=1.25in,clip,keepaspectratio]{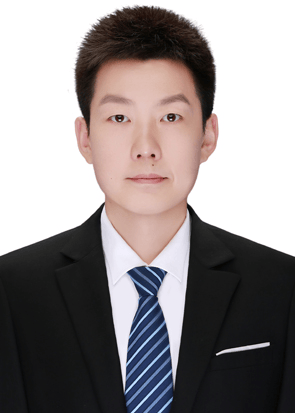}}]{Geng Sun} (Senior Member, IEEE) received the B.S. degree in communication engineering from Dalian Polytechnic University, and the Ph.D. degree in computer science and technology from Jilin University, in 2011 and 2018, respectively. He was a Visiting Researcher with the School of Electrical and Computer Engineering, Georgia Institute of Technology, USA. He is a Professor in the College of Computer Science and Technology at Jilin University. Currently, he is working as a visiting scholar at the College of Computing and Data Science, Nanyang Technological University, Singapore. He has published over 100 high-quality papers, including IEEE TMC, IEEE JSAC, IEEE/ACM ToN, IEEE TWC, IEEE TCOM, IEEE TAP, IEEE IoT-J, IEEE TIM, IEEE INFOCOM, IEEE GLOBECOM, and IEEE ICC. He serves as the Associate Editors of IEEE Communications Surveys \& Tutorials, IEEE Transactions on Communications, IEEE Transactions on Vehicular Technology, IEEE Transactions on Network Science and Engineering, IEEE Transactions on Network and Service Management and IEEE Networking Letters. He serves as the Lead Guest Editor of Special Issues for IEEE Transactions on Network Science and Engineering, IEEE Internet of Things Journal, IEEE Networking Letters. He also serves as the Guest Editor of Special Issues for IEEE Transactions on Services Computing, IEEE Communications Magazine, and IEEE Open Journal of the Communications Society. His research interests include Low-altitude Wireless Networks, UAV communications and Networking, Mobile Edge Computing (MEC), Intelligent Reflecting Surface (IRS), Generative AI and Agentic AI, and deep reinforcement learning.
\end{IEEEbiography}
\vspace{-10 mm}

\begin{IEEEbiography}[{\includegraphics[width=1in,height=1.25in,clip,keepaspectratio]{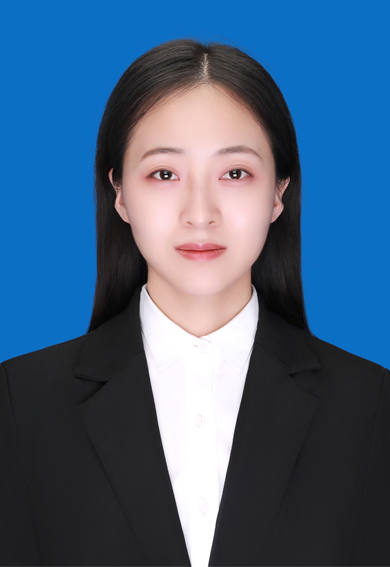}}]{Zemin Sun}
(Member, IEEE) received a BS degree in Software Engineering, an MS degree and a Ph.D degree in Computer Science and Technology from Jilin University, Changchun, China, in 2015, 2018, and 2022, respectively. She was a visiting Ph.D. student at the University of Waterloo. She currently serves as an assistant researcher in the College of Computer Science and Technology at Jilin University. Her research interests include vehicular networks, edge computing, and game theory.
\end{IEEEbiography}
\vspace{-40 mm}

\begin{IEEEbiography}[{\includegraphics[width=1in,height=1.25in,clip,keepaspectratio]{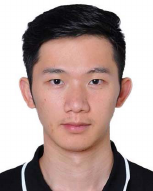}}]{Jiacheng Wang} received the Ph.D. degree from the School of Communication and Information Engineering, Chongqing University of Posts and Telecommunications, Chongqing, China. He is currently a Research Associate in computer science and engineering with Nanyang Technological University, Singapore. His research interests include wireless sensing, semantic communications, and metaverse.
\end{IEEEbiography}
\vspace{-40 mm}

\begin{IEEEbiography}[{\includegraphics[width=1in,height=1.25in,clip,keepaspectratio]{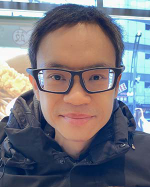}}]{Dusit Niyato} (Fellow, IEEE) is a professor in the College of Computing and Data Science, at Nanyang Technological University, Singapore. He received B.Eng. from King Mongkuts Institute of Technology Ladkrabang (KMITL), Thailand and Ph.D. in Electrical and Computer Engineering from the University of Manitoba, Canada. His research interests are in the areas of mobile generative AI, edge intelligence, decentralized machine learning, and incentive mechanism design.
\end{IEEEbiography}
\vspace{-40 mm}

\begin{IEEEbiography}[{\includegraphics[width=1in,height=1.25in,clip,keepaspectratio]{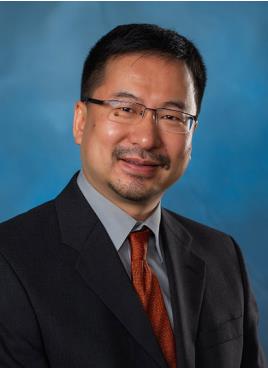}}]{Shiwen Mao} (Fellow, IEEE) is a Professor and  the Earle C. Williams Eminent Scholar Chair,  and the Director of the Wireless Engineering Research and Education Center, Auburn University,  Auburn, AL, USA. His research interest includes  wireless networks, multimedia communications,  and smart grid. He received the IEEE ComSoc  MMTC Outstanding Researcher Award in 2023,  the IEEE ComSoc TC-CSR Distinguished Technical Achievement Award in 2019, and the NSF  CAREER Award in 2010. He is a co-recipient of  the 2022 Best Journal Paper Award of IEEE ComSoc eHealth Technical Committee, the 2021 Best Paper Award of Elsevier/Digital Communications and Networks (KeAi), the 2021 IEEE Internet of Things  Journal Best Paper Award, the 2021 IEEE Communications Society  Outstanding Paper Award, the IEEE Vehicular Technology Society 2020  Jack Neubauer Memorial Award, the 2018 ComSoc MMTC Best Journal  Paper Award and the 2017 Best Conference Paper Award, the 2004  IEEE Communications Society Leonard G. Abraham Prize in the Field  of Communications Systems, and several ComSoc technical committee  and conference best paper/demo awards. He is the Editor-in-Chief  of IEEE TRANSACTIONS ON COGNITIVE COMMUNICATIONS AND  NETWORKING. He is a Distinguished Lecturer of IEEE Communications Society and the IEEE Council of RFID.
\end{IEEEbiography}

% \vfill
\end{document}

\title{Supplemental Material \\
\begin{large} 
   \textit{Low-Altitude Satellite-AAV Collaborative Joint Mobile Edge Computing and Data Collection via Diffusion-based Deep Reinforcement Learning}
\end{large} }

\maketitle

\appendices

{
%section
%Large Language Model (LLM)-based Operating Point Selection
%

\section{Discussion about Alternative Multi-objective Approaches}

\par This part discusses the applicability and challenges of alternative multi-objective optimization methods compared to the weighted sum approach, such as Pareto optimality, hierarchical optimization, and dynamic weight adjustment.

\par The Pareto optimality framework offers a theoretically rigorous alternative for multi-objective optimization, while posing significant computational and practical challenges. Specifically, this approach does not predefine weights but instead computes a set of Pareto optimal solutions, where any improvement in one objective necessarily leads to a deterioration in at least one other. Within this framework, multi-objective DRL algorithms can generate a complete Pareto frontier, providing decision-makers with a comprehensive perspective on trade-offs and granting them the flexibility to select optimal strategies online. However, multi-objective DRL significantly increases computational overhead and training complexity. Furthermore, in our proposed QAGOB method, which generates Q-value-based actions through a diffusion model, the core challenge remains how to integrate multiple potentially conflicting Q-value signals into the denoising process of the diffusion model.

\par Hierarchical optimization approaches establish strict priority rankings among objectives, which prove effective for scenarios with rigid quality-of-service requirements but may be poorly suited for problems with deeply coupled objectives. For instance, this methodology would prioritize maintaining all tasks within their maximum tolerable delay thresholds, subsequently optimize DC volume subject to these temporal constraints, and finally address energy minimization within the remaining feasible space. Such hierarchical structuring proves particularly effective in scenarios with strict quality-of-service requirements, such as time-sensitive tasks like fire warning systems.

\par However, in our joint optimization problem, the objectives exhibit deep coupling. Specifically, reducing latency may require the AAV to fly closer to the target, while collecting more data extends hover time, thereby increasing energy consumption. Consequently, strict hierarchical separation may yield suboptimal solutions that fail to fully leverage the trade-offs between objectives. Furthermore, converting one objective into a hard constraint sharply reduces the feasible solution space, potentially making it difficult for the algorithm to find viable solutions.

\par Instead of the fixed weights used in this paper, dynamic weight tuning methods can rebalance the three objectives in real time. For instance, the energy weight coefficient can be increased when the AAV battery is low, or the delay penalty coefficient can be raised during urgent missions. This mechanism offers greater flexibility than static weights. However, designing dynamic weights inherently poses a complex control or optimization problem. It requires defining explicit adjustment rules or learning a weight adjustment strategy, which increases system complexity and introduces additional hyperparameters.

\par In summary, although other multi-objective approaches like Pareto optimality are theoretically more comprehensive, weighted summation is widely adopted in practical engineering applications due to its simplicity, interpretability, and ease of implementation. For the proposed diffusion-based DRL framework, which already exhibits high computational complexity, employing weighted summation is a pragmatic choice to ensure the trainability and convergence of the algorithm.

\section{Algorithm Details}

\par In this section, we present some algorithmic details about the network architecture of the actor-critic component. As can be seen from Fig. 2 in the main text, the proposed QAGOB method mainly consists of a diffusion-based actor network, two critic networks, the target networks of actors and critics, and two replay buffers. The specific details are as follows:

\begin{itemize}
    \item \textit{Diffusion-based Actor Network:} In the QAGOB method, the core of the actor-network is a diffusion model, rather than a conventional multi-layer perceptron (MLP). At each time step $t$, the state $s_t$ is encoded as a state vector and injected into multiple layers of the internal network of the diffusion model as conditional input. Through an iterative denoising process, it generates the final action from a fully random noise vector. This diffusion process captures dependencies from state to action and explores more complex and diverse multimodal action distributions compared to a single forward pass through an MLP, thereby generating high-quality actions.

    \item \textit{Twin Critic Network:} During the policy improvement process, a double critic network is employed to reduce excessive estimation bias. Each critic network adopts a MLP architecture and independently updates its own parameters using the same optimization target. During the training process, the minimum Q-value estimate from the two critic networks is utilized.

    \item \textit{Target Networks:} Each critic network and actor network has its own corresponding target network, which has an identical architecture to its online network. However, the parameters of the target network are not updated directly via gradient descent. Instead, they are slowly copied from the online network through a soft update mechanism, providing a slowly changing and relatively stable optimization target. This significantly enhances the stability of the training process.
    
    \item \textit{Replay Buffer and Diffusion Buffer:} The QAGOB approach employs two types of replay buffers. On the one hand, the main replay buffer stores transitions $\{s_t,a_t,r_t,s_{t+1}\}$ obtained through interactions with the environment. When updating the critic network, a batch of transitions is randomly sampled from this buffer to compute the temporal difference target and the mean squared error of the critic network. On the other hand, the diffusion buffer is specifically designed for the diffusion actor network, which stores high-quality state-action pairs to enhance the training efficiency of the diffusion actor network.
\end{itemize}

% \section{Performance Analysis}

% \par In this work, we further provide some additional simulation results about the energy management of the satellite and GDs, as well as the comparison results between the joint optimization and separate optimization of MEC and DC.

\begin{figure}[tbp]
    \centering
    \includegraphics[width=\linewidth]{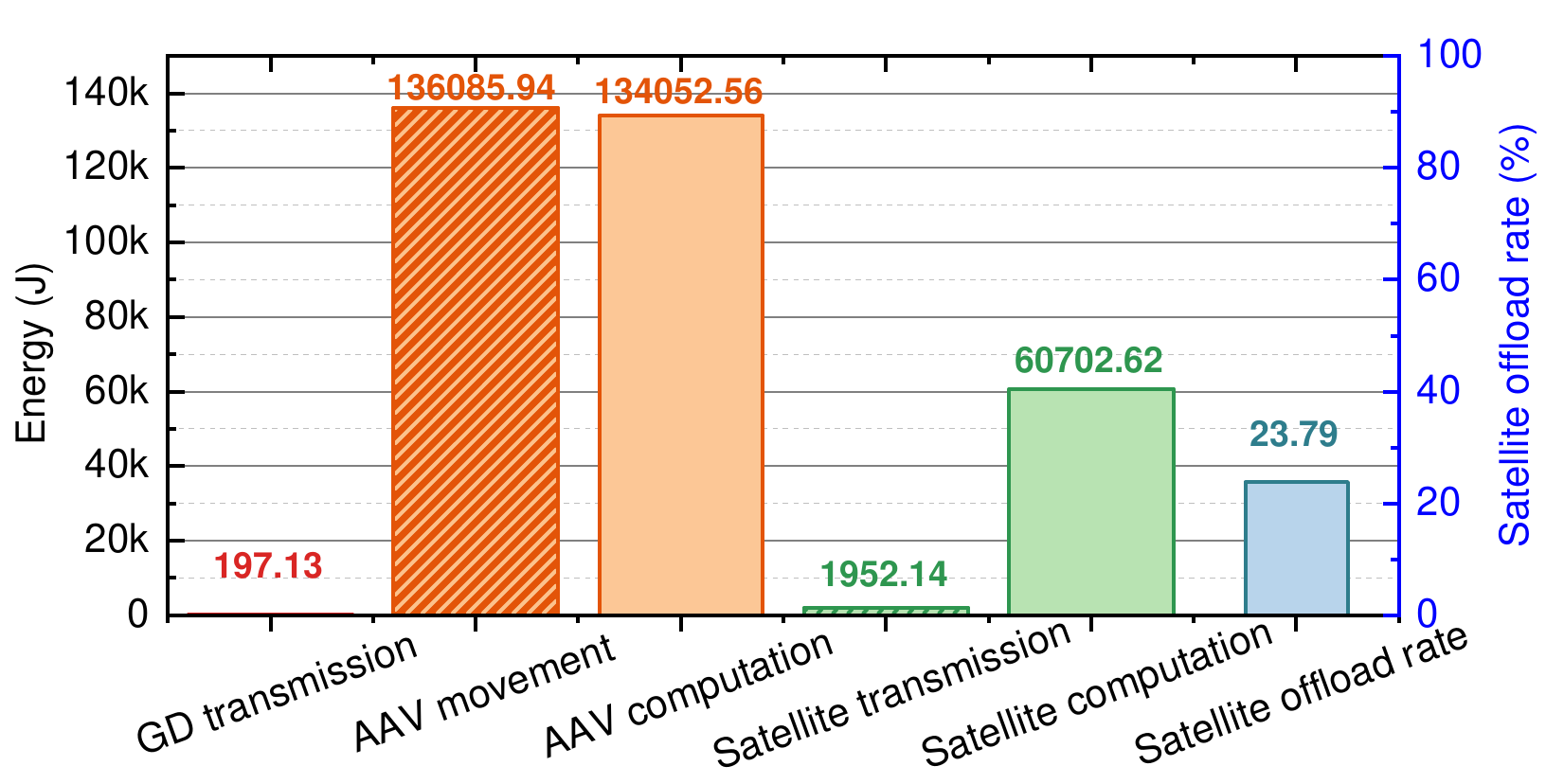}
    \caption{Satellite offload ratio and energy consumption of GD transmission, AAV movement, AAV computation, satellite transmission, and satellite computation.}
    \label{fig:all energy}
\end{figure}

\section{Performance Analysis}

\par In this work, we further provide some additional simulation results about the energy management of the satellite and GDs, as well as the comparison results between the joint optimization and separate optimization of MEC and DC.

\subsection{Energy Analysis of The Satellite and GDs}

\par In this part, we discuss the energy consumption of the satellite and GDs.

\par Fig. \ref{fig:all energy} illustrate the energy consumption of all the components in the considered system, as well as the offload rate of the MEC tasks to the satellite. As can be seen, the energy consumed by GDs during transmission is far less than that of the AAVs and the satellite, and the energy consumption of the satellite accounts for less than 20\% of the total. In contrast, the movement and computation energy consumption of the AAVs dominate the system. Additionally, we observed that only 23.79\% of MEC tasks are further offloaded to the satellites, indicating that the system strategy tends to process tasks on the AAVs in order to reduce E2E latency. Therefore, the energy consumption of the AAVs is the primary decision-making factor in the considered system, which affects the performance bottleneck of the system.

\begin{figure}[tp]
    \centering
    \includegraphics[width=\linewidth]{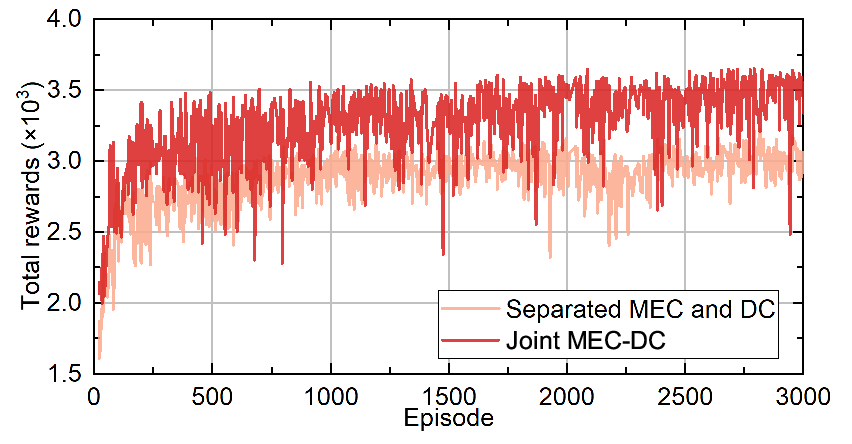}
    \caption{Reward curves of joint MEC-DC and separated MEC-DC framework.}
    \label{fig:separate converge}
\end{figure}

\subsection{Comparison with The Separated MEC and DC}

\begin{figure}[tp]
    \centering
    \includegraphics[width=\linewidth]{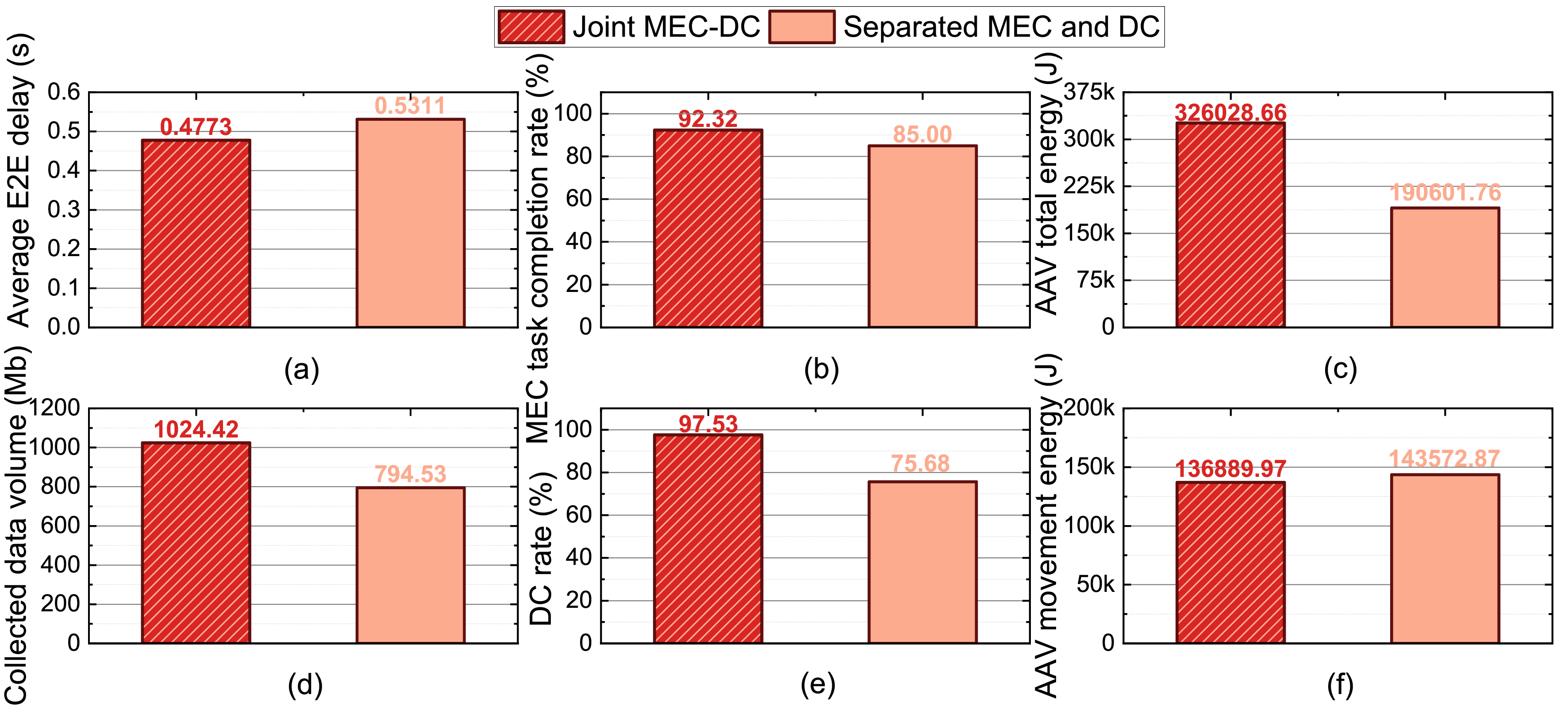}
    \caption{Comparison results of joint MEC-DC and separated MEC-DC framework.}
    \label{fig:separate metric}
\end{figure}

\begin{figure}[tp]
    \centering
    \subfloat[\label{fig:a}]{
    \includegraphics[width=0.49\linewidth]{figure/joint_tra.pdf}}
    \subfloat[\label{fig:b}]{
    \includegraphics[width=0.49\linewidth]{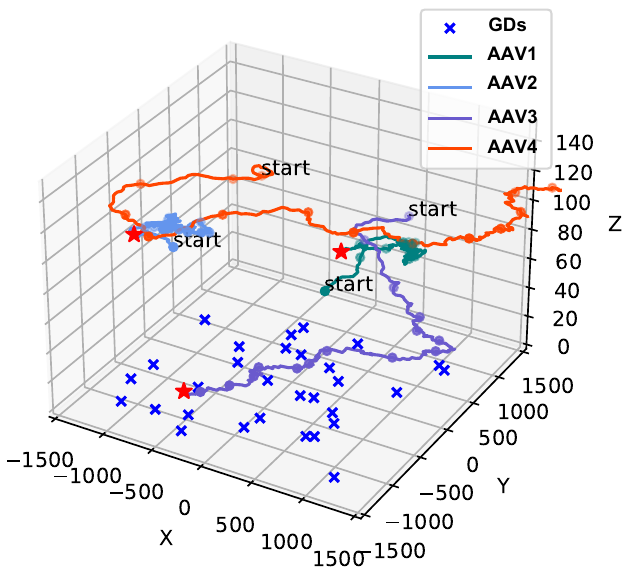}}
    \caption{The AAV movement trajectory, the start and end points are marked with text and red stars, respectively. (a) Joint MEC-DC optimization framework. (b) Separate MEC-DC optimization framework. (AAV1 and AAV2 are MEC-AAV, AAV3 and AAV4 are DC-AAV).}
    \label{fig:twotrajectory}
\end{figure}

\par This part compares the performance of the joint MEC-DC optimization with the separated MEC and DC optimization, including the convergence results, comparison results, and trajectory results, respectively.

\par Fig. \ref{fig:separate converge} shows the training rewards of the QAGOB method under the joint MEC-DC framework and the separate MEC and DC framework. It can be observed that the reward level achieved through the joint optimization is significantly higher than that of the separate optimization baseline. This is likely because the optimization variables (such as AAV trajectory planning, resource allocation, energy management, and device association) are coupled between the MEC and DC systems, which means that optimizing one objective separately may degrade the other objective, leading to suboptimal solutions. In contrast, joint optimization takes both objectives into account and learns a strategy that achieves an appropriate trade-off, thereby maximizing the overall utility of the system.

\par Fig. \ref{fig:separate metric} exhibits the system performance under joint and separate frameworks, including average E2E latency, MEC task completion rate, AAV energy consumption, amount of collected data, DC task completion rate, and AAV movement energy consumption. As shown in the figure, the joint optimization framework outperforms separate optimization in both E2E latency and data collection amount. While maintaining higher task completion rates, joint optimization also achieves lower movement energy consumption. This suggests that joint optimization captures conflicts and dependencies between objectives to find globally optimal trade-off strategies. Additionally, AAVs can collect data during MEC idle periods, maximizing resource utilization efficiency.

\par Figs. \ref{fig:a} and \ref{fig:b} present the visualized trajectories of AAVs under the joint MEC-DC optimization framework and the separate optimization framework, respectively. It can be observed that the two frameworks guide the AAVs to adopt distinctly different mobility strategies. Under the joint optimization framework, all AAVs execute moderate short-distance movements across a wide area, effectively balancing the requirement of MEC for low-latency service with the requirement of DC for broad data coverage. In contrast, under the separate optimization framework, MEC-AAVs tend to make minor positional adjustments within localized small areas to ensure the real-time performance of computation tasks, while DC-AAVs are required to fly over long distances across large areas in order to maximize DC. Although each type of AAV performs optimally with respect to its individual objective, the strategies are disjointed and lack synergistic coordination, resulting in fragmented resource utilization and overall inefficiency.

\par In summary, compared to the separate optimization of MEC and DC, the joint optimization better handles the conflicts and interdependencies between the two objectives. It achieves synergistic coordination between MEC and DC objectives at the strategic level and significantly enhances overall service efficiency at the system level.

% \section{Channel Estimation Methods}

% \bibliographystyle{IEEEtran}
% \bibliography{bib.bib}